\documentclass[pageno]{jpaper}

\usepackage[normalem]{ulem}
\usepackage{tikz}
\usepackage{amsmath}
\usepackage{algorithm}
\usepackage{algorithmic}
\usepackage{balance}
\usepackage{hyperref}
\usepackage{comment}
\usepackage{multirow}
\usepackage[symbol]{footmisc}
\usepackage{bbding}
\usepackage{array}

\newcommand{\tabincell}[2]{\begin{tabular}[t]{@{}#1@{}}#2\end{tabular}}

\newcommand{\egg}[1] {}

\begin{document}


\title{\textit{Mimose}: An Input-Aware Checkpointing Planner for Efficient Training on GPU}

\author{Jianjin Liao$^{1\dag}$, Mingzhen Li$^{1\dag}$, Qingxiao Sun$^1$, Jiwei Hao$^1$, Fengwei Yu$^2$, Shengdong Chen$^2$ \\[.3em]Ye Tao$^2$, Zicheng Zhang$^2$, Hailong Yang$^{1*}$, Zhongzhi Luan$^1$, Depei Qian$^1$ \\[.3em]
\textit{Beihang University$^1$  \quad SenseTime Research$^2$ } \\[.3em]
}

\date{}
\maketitle

\newcommand\blfootnote[1]{%
	\begingroup
	\renewcommand\thefootnote{}\footnote{#1}%
	\addtocounter{footnote}{-1}%
	\endgroup
}
\blfootnote{Email: \{liaojianjin, lmzhhh, qingxiaosun, biaosun, hailong.yang, 07680, depeiq\}@buaa.edu.cn, \{yufengwei, chenshengdong, taoye1, zhangzicheng\}@sensetime.com}
\blfootnote{$\dag$ Contributed equally.}
\blfootnote{$*$ Corresponding author.}

\thispagestyle{empty}

\begin{abstract}

Larger deep learning models usually lead to higher model quality with an ever-increasing GPU memory footprint. Although tensor checkpointing techniques have been proposed to enable training under a restricted GPU memory budget, the input tensor dynamics have been unexploited for optimizing performance while reducing GPU memory footprint. Specifically, due to the diverse datasets and subsequent data argumentation, the input tensor size per mini-batch is dynamic during the training process, leading to a changing GPU memory footprint. However, to leverage such input tensor dynamics in checkpointing, there are two challenges to be solved. First, the checkpointing plan needs to be determined during runtime due to the dynamics of input tensors. Second, the checkpointing plan needs to be applied on the fly without significantly deteriorating the performance.

In this paper, we propose \textit{Mimose}, an input-aware tensor checkpointing planner respecting the memory budget while enabling efficient model training on GPU. 
\textit{Mimose} builds a lightweight but accurate prediction model of GPU memory usage online, without pre-analyzing the model. It generates a tensor checkpointing plan based on per-layer memory prediction and applies it to training progress on the fly. It also adopts a caching strategy to avoid having to regenerate the plan for repeated input size.
Our experiments show that \textit{Mimose} achieves superior training throughput compared to state-of-the-art memory planners under the same GPU memory budgets.
	
\end{abstract}
\section{Introduction}
\label{sec:introduction}

Deep learning (DL) models are important and indispensable building blocks of various fields, such as natural language processing~\cite{ floridi2020gpt,raffel2020exploring,zaheer2020big}, object detection~\cite{girshick2015fast,redmon2018yolov3}, and autonomous driving~\cite{9283940,zhou2018voxelnet}.
DL models are prevalently becoming larger to achieve higher quality, with such trend expected to continue~\cite{ghorbani2022scaling, https://doi.org/10.48550/arxiv.2001.08361}.
Training large models necessitate an ever-increasing GPU memory footprint. However, the GPU memory capacity grows at a slower rate~\cite{zero-infinity, capuchin, checkmate}. The unsatisfied demand for training large models due to limited GPU memory prevents DL practitioners from experimenting and innovating cutting-edge models, thereby impeding the rapid advance of the DL field.
As reported by previous works~\cite{zero-infinity,capuchin,checkmate,chen2016training}, the GPU memory usage during model training is dominated by the intermediate activation tensors (activations in short).

Activations are intermediate outputs generated by DL operators in the forward pass, and then kept in GPU memory until consumed to calculate the gradients in the backward pass. To reduce the memory occupancy of activations, a large number of techniques have been proposed. These techniques can be classified into three categories: compressing~\cite{courbariaux2014training,gupta2015deep,judd2016proteus,jain2018gist,han2015deep,hubara2016binarized,han2015learning,han2016eie,yang2021auto}, swapping~\cite{rhu2016vdnn,meng2017training,shriram2019dynamic,huang2020swapadvisor,ren2021sentinel}, and checkpointing~\cite{chen2016training,checkmate,dtr,feng2021optimal}. 
Compressing attempt to convert activation into its low-bit counterpart, and thus may affect the convergence and the model quality, due to the iterative nature of training may lead to uncontrollable error propagation. 
Swapping offloads the activations from GPU memory to CPU DRAM in the forward pass, and asynchronously copies them back in the backward pass. Unfortunately, the copying overhead is quite high due to the limited PCIe bandwidth. Checkpointing allows dropping the activations in the forward pass and re-generating them by replaying the forward computation (i.e., re-computation) in the backward pass. In general, due to the lower overhead of re-computation than data transmission between GPU and CPU, checkpointing is widely adopted by popular DL frameworks such as TensorFlow~\cite{checkmate} and PyTorch~\cite{dtr, feng2021optimal} to reduce GPU memory consumption during model training.

The fundamental of a checkpointing technique is the GPU memory planner that decides when and where to drop and re-compute the activations. Depending on whether it requires prior knowledge of the model structure, the GPU memory planner can be further divided into static planners (e.g., Checkmate~\cite{checkmate}) and dynamic planners (e.g., DTR~\cite{dtr}). The static planners commonly adopt a conservative plan respecting the largest input tensor to avoid GPU memory over-subscription. Whereas, the dynamic planners reactively checkpoint the activations with the lowest re-computing costs in a greedy manner, when running out of GPU memory. However, both types of planners fail to consider input dynamics during training, thus unnecessarily sacrificing the training performance for reduced GPU memory occupancy. For example, even if the GPU memory is sufficient when the input size is small, the conservation of static planners results in massive redundant computation. Whereas, due to the lack of model structure knowledge, dynamic planners may re-generate the checkpointing plan for the same input size redundantly, and thus deteriorate the training performance (details in Section~\ref{sec:motivation_2}). 

The input dynamics with changing activation sizes exist during training due to the diverse datasets and subsequent data argumentation, which results in changing GPU memory footprint (details in Section~\ref{sec:motivation_1}).
For example, object detection datasets (e.g., COCO) for computer vision tasks contain many images with varying aspect ratios, whereas NLP (e.g., SWAG) datasets contain many multiple-choice questions with varying text sequence lengths.
As for subsequent data argumentation, a image can be resized to a random size while keeping its aspect ratio unchanged~\cite{Detectron2018}, to increase the model robustness. At the same time, a text sequence can be broken down into a sequence of word tokens during the tokenization.
Besides, several images/texts are then collated into a mini-batch after padding and truncation. However, such input dynamics can hardly be exploited by current GPU memory planners. To leverage the input dynamics in checkpointing, two challenges must be solved. First, the checkpointing plan needs to be determined during runtime due to the input dynamics. Second, the checkpointing plan needs to be applied on the fly without significantly deteriorating the performance.

In many production scenarios, such as personalized recommendation and meta universe, the DL models need to be frequently finetuned to fit the latest collected dataset to avoid the so-called "concept drift"~\cite{concept-drift} and improve the serving quality of the model. 
Therefore, input tensor dynamics become more severe. It is unrealistic to obtain the size distribution of the dataset and perform checkpointing plan in advance, as the input size is unpredictable.
Therefore, in the whole training pipeline, a checkpointing planner has no prior knowledge of the dataset distribution, data argumentation process, model structure, and model parameters, and thus cannot predict the memory usage of the model. 
Therefore, the checkpointing planner should address the following challenges.
\textit{1)} It should collect the GPU memory usage online without prior knowledge.
\textit{2)} It should predict the per-layer GPU memory usage quickly and accurately, given arbitrary input tensors.
\textit{3)} It should generate and apply checkpointing plans agilely during runtime based on previous memory prediction.
Putting the above together, its overhead should be trivial to affect the training efficiency.

In this work, we propose \textit{Mimose}, an input-aware checkpointing planner respecting the memory budget while enabling efficient model training on GPU. 
The key feature of \textit{Mimose} is that it dynamically adjusts the checkpointing plan according to the predicted memory usage of current input tensor, in order to maximize GPU memory utilization and minimize the performance overhead.
\textit{Mimose} builds a lightweight but accurate prediction model of GPU memory usage online without pre-analyzing the model, to achieve the sub-millisecond-level checkpointing planning for each input tensor. It generates a tensor checkpointing plan and applies the plan on the fly during the training progress. 
The key contributions of the paper are as follows.
\begin{itemize}
	\item We propose an online GPU memory estimator that predicts the memory usage of activation tensors for the given input size. The estimator is constructed during model training without prior knowledge of the model structure or the input size. After negligible training iterations, the estimator can offer accurate enough memory prediction to facilitate generating checkpointing plans.
	\item We propose an effective checkpointing scheduler that generates and applies the checkpointing plans based on the memory estimation during runtime. In addition, it adopts a caching strategy to avoid re-generating the checkpointing plans for repeated input sizes redundantly.
	\item We develop the \textit{Mimose} framework with the input-aware checkpointing planner for efficient training on GPU. \textit{Mimose} works entirely online, and does not rely on either model pre-analyzing or ahead-of-time memory planning. Our experiment results demonstrate that \textit{Mimose} can achieve better training performance than state-of-the-art checkpointing techniques.
\end{itemize}

The remaining part of this paper is organized as follows. Section~\ref{sec:background} describes the background of the DL training pipeline as well as prior checkpointing planners. Section~\ref{sec:motivation} presents the input tensor dynamics and drawbacks of prior checkpointing planners, which motivates \textit{Mimose}. Section~\ref{sec:design} and Section~\ref{sec:implementation} present the methodology and implementation of this input-aware checkpointing planner. Section~\ref{sec:evaluation} presents the evaluation results of \textit{Mimose} by comparing to the state-of-the-art memory planners. Section~\ref{sec:relatedwork} discusses the related work, and Section~\ref{sec:conclusion} concludes this paper.

\section{Background}
\label{sec:background}

\subsection{Training Pipeline}

\begin{figure}[htbp]
	\centering
	\includegraphics[width=\linewidth]{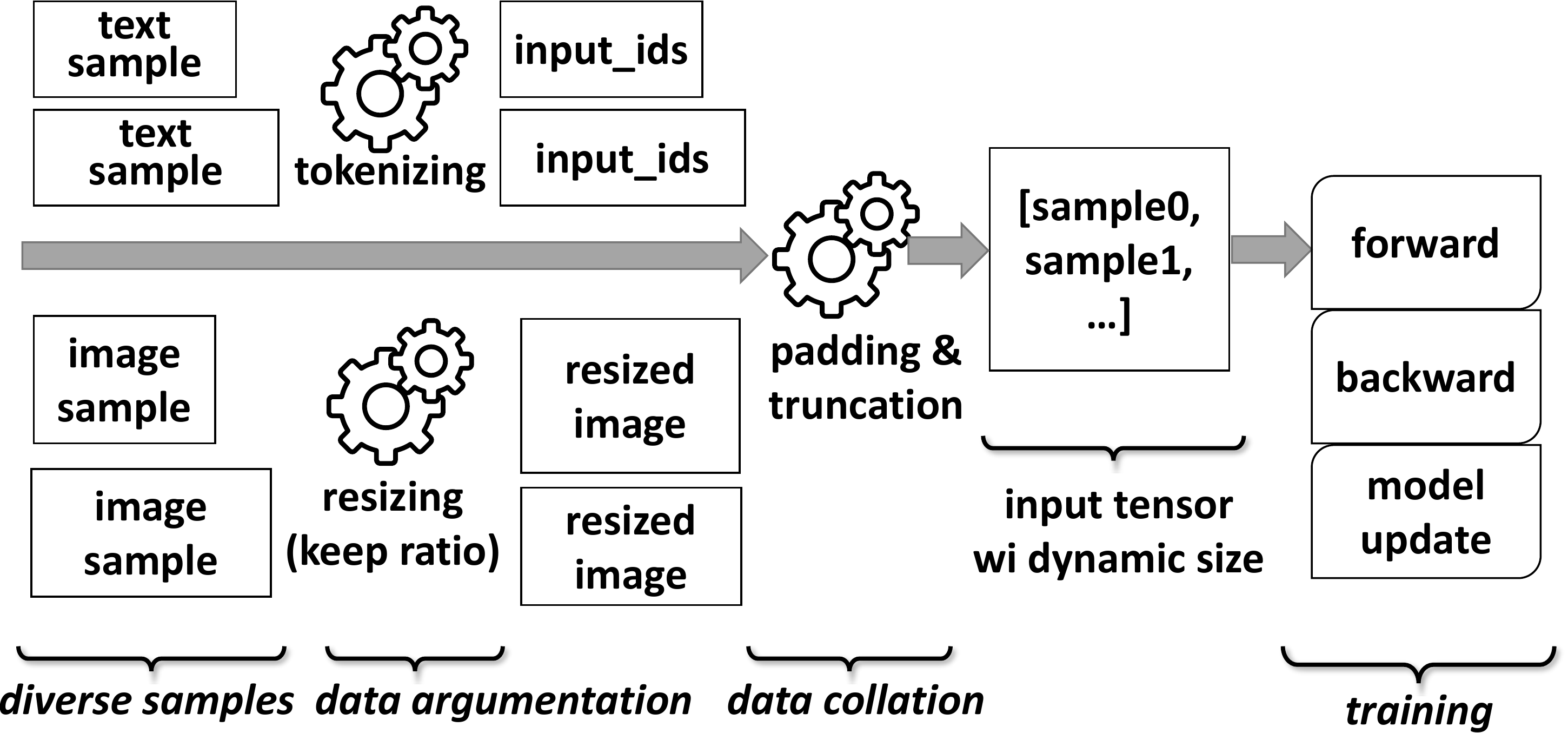}
	\caption{Input tensor dynamics in the training pipeline, which is mainly caused by the dataset and the data argumentation.}
	\label{fig:training-pipeline}
\end{figure}

Deep learning training often includes many iterations, with each iteration processing a mini-batch containing a few samples.
In each iteration, samples are processed in four typical phases and finally update the model parameters, as shown in Figure~\ref{fig:training-pipeline}. 
First, samples are loaded from the training dataset and preprocessed through data argumentation and collation to form a mini-batch input tensor. 
For instance, text samples of question answering are collected from various sources with diverse text lengths. After tokenizing, samples are split as tokens and converted into sequences (e.g., \textit{input\_ids}), which are not always of the same length across samples. 
Shorter sequences in a mini-batch are padded to match the longest sequence, and the sequences too long to handle are truncated shorter. The sequences with uniform shapes are then collated to an input tensor. Note that the input tensor sizes can fluctuate across iterations due to the diversity of datasets and the flexibility of data argumentation.
Second, the input tensor is used for calculation in conjunction with the model parameters to produce a set of scores, a process known as the \textit{forward pass}. During the forward pass, the activation tensors are generated successively and stored in GPU memory for reuse in the backward. At the end of the forward pass, a train loss is derived by comparing the produced scores with the desired scores.
Third, the loss is propagated through the entire model to compute the gradients of the model parameters, a process known as the \textit{backward pass}. During the backward pass, the activation tensors are deallocated immediately once corresponding gradients are derived.
Finally, the gradients are scaled by a learning rate and used to \textit{update} the model parameters.

\subsection{Checkpointing Planners}

\begin{figure}[htbp]
	\centering
	\includegraphics[width=\linewidth]{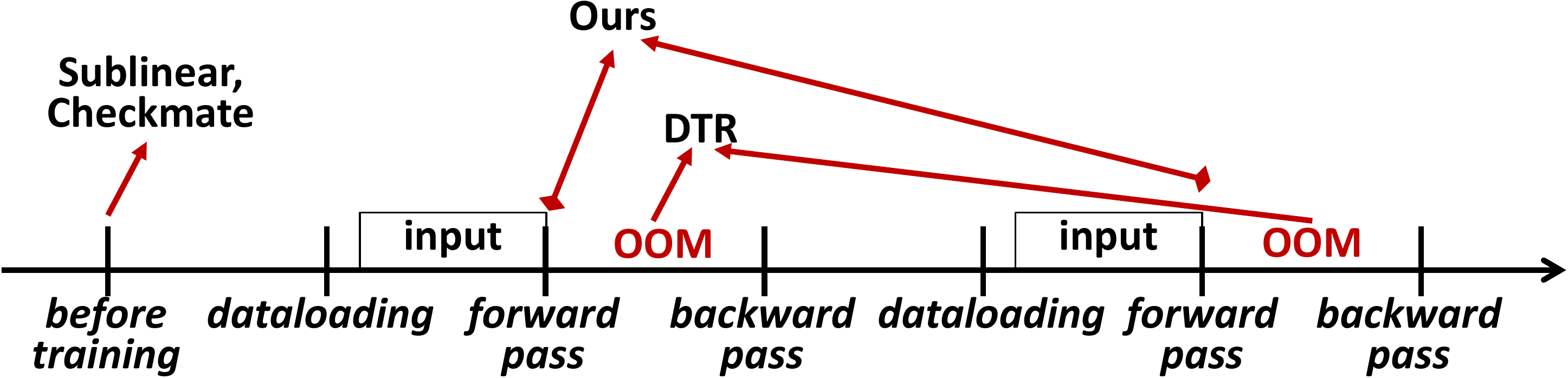}
	\caption{Comparison across prior checkpointing planners, where x-axis indicates the timeline. The red arrow indicates the moment of checkpointing plan generation.}
	\label{fig:training-chekpointing}
\end{figure}

Mainstream memory planners reduce GPU memory footprint by techniques including compressing, swapping, or checkpointing. Compressing can have a significant impact on convergence speed and model accuracy~\cite{huang2020swapadvisor}. The time cost incurred by swapping is more than 2$\times$ the computation time for most layers, which may slow down the training process~\cite{he2022home}. Due to the above considerations, we leverage checkpointing for efficient model training on GPU while respecting the memory budget. Figure~\ref{fig:training-chekpointing} shows the comparison across checkpointing planners, where the red arrow indicates the moment of checkpointing plan generation.

Static planners~\cite{chen2016training,checkmate} collect model information and generate the checkpointing plan before training. \textit{Sublinear}~\cite{chen2016training} reduces the memory cost to store the feature maps through computation graph liveness analysis. \textit{Checkmate}~\cite{checkmate} searches near-optimal checkpointing plans using mixed an integer linear program (MILP) solver with an approximation algorithm. However, static planners cannot effectively handle tasks with input dynamics. To avoid exceeding the memory capacity during training, static planners have to generate checkpointing plans based on a predefined maximum-sized input. Such checkpointing plans sacrifice training speed due to additional recomputation overhead for small-sized inputs.

In contrast, the dynamic planner (i.e., DTR~\cite{dtr}) can handle input dynamics by generating checkpointing plans on the fly. \textit{DTR} dynamically gathers information about tensors and operators and greedily drops activations when an OOM exception happens. However, the dropping operations of \textit{DTR} are triggered on demand by the OOM exception, which increases the checkpointing delay compared to static plans. Furthermore, \textit{DTR} lacks holistic information about model training and ignores historically generated plans. This makes \textit{DTR} regenerate plans for duplicate input sizes with significant overhead.

From the above analysis, it is necessary to design an input-aware memory planner that can reuse checkpointing plans of the same input sizes. In addition, lightweight regression algorithms can be utilized to predict memory usage based on current input without pre-analyzing the model.

\section{Motivation}
\label{sec:motivation}

\begin{figure*}[htbp]
	\begin{minipage}[t]{0.34\linewidth}
		\centering 
		\includegraphics[width=\linewidth]{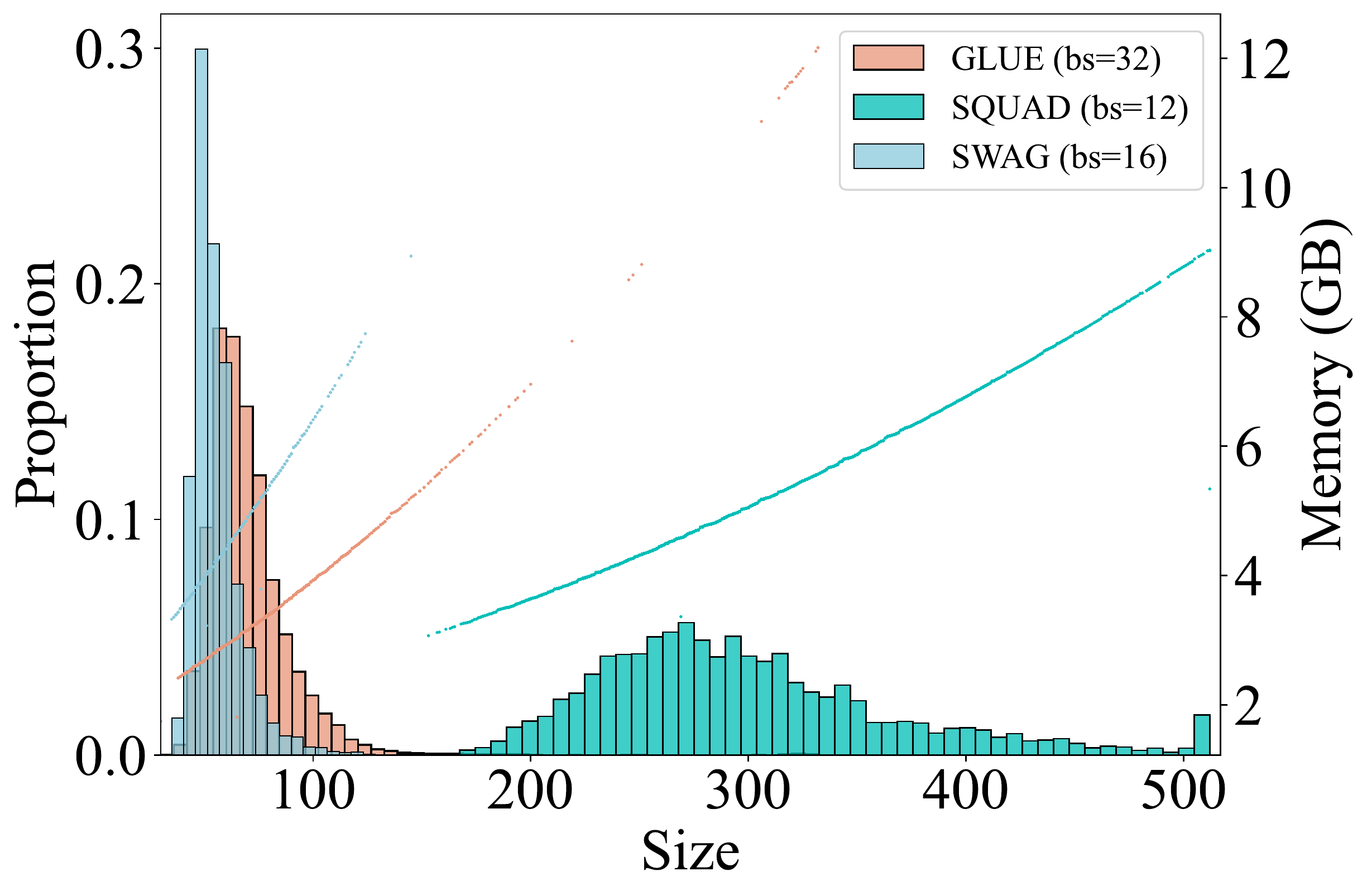} 
		\caption{Input size distributions of SWAG, SQuAD, GLUE-QQP datasets (left y-axis) and GPU memory footprints (right y-axis) when training Bert-base with batch size set to 16, 12, 32, respectively.}
		\label{fig:dataset-nlp}
	\end{minipage}
	\hfill
	\begin{minipage}[t]{0.31\linewidth}
		\centering 
		\includegraphics[width=0.97\linewidth]{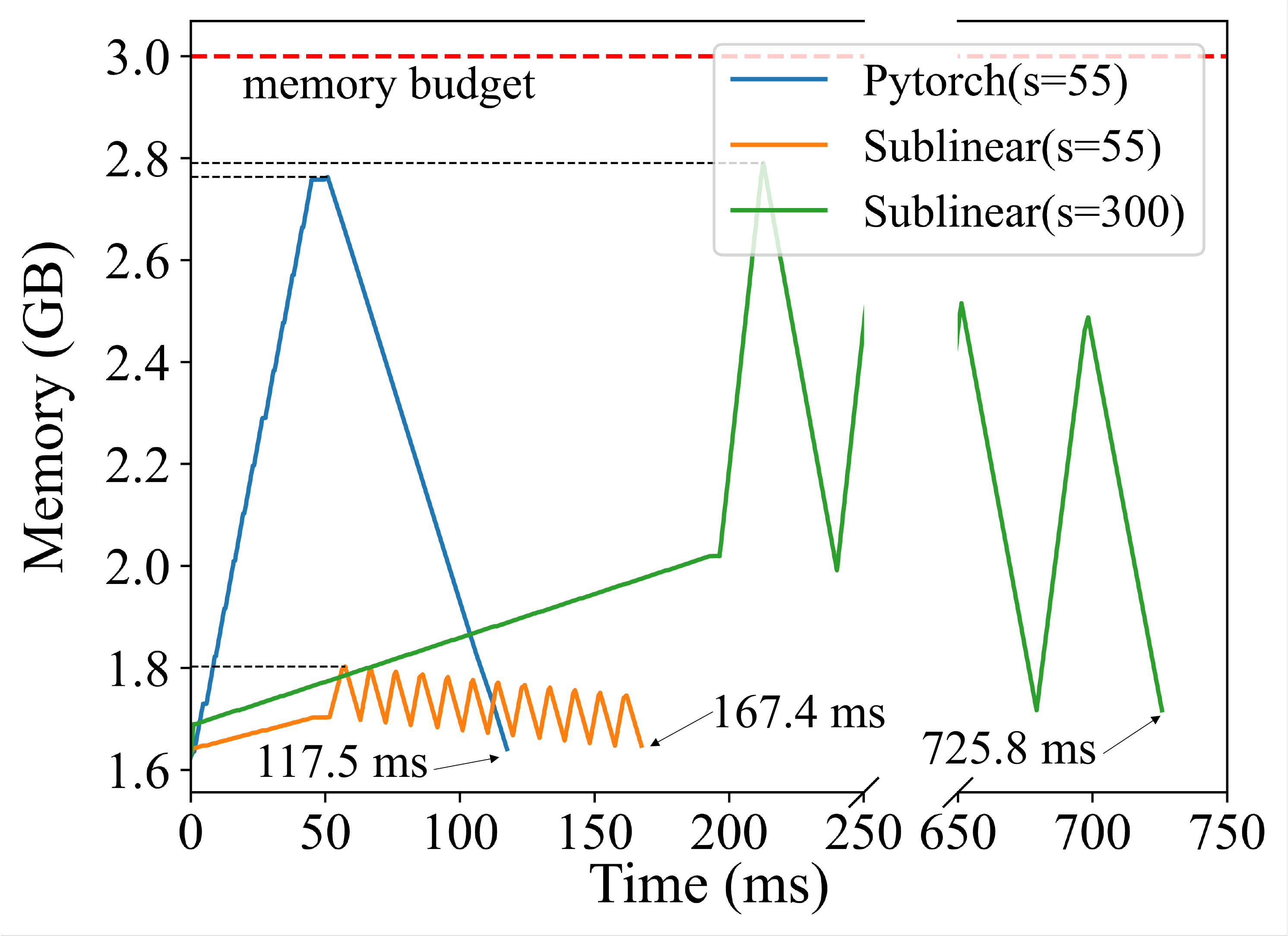} 
		\caption{Memory footprint curves of training Bert-base on GLUE-QQP dataset using static checkpointing planner, \textit{sublinear}. The memory budget is set to 3 GB.}
		\label{fig:small-input-size-gap}
	\end{minipage}
	\hfill
	\begin{minipage}[t]{0.32\linewidth} 
		\centering 
		\includegraphics[width=0.92\linewidth]{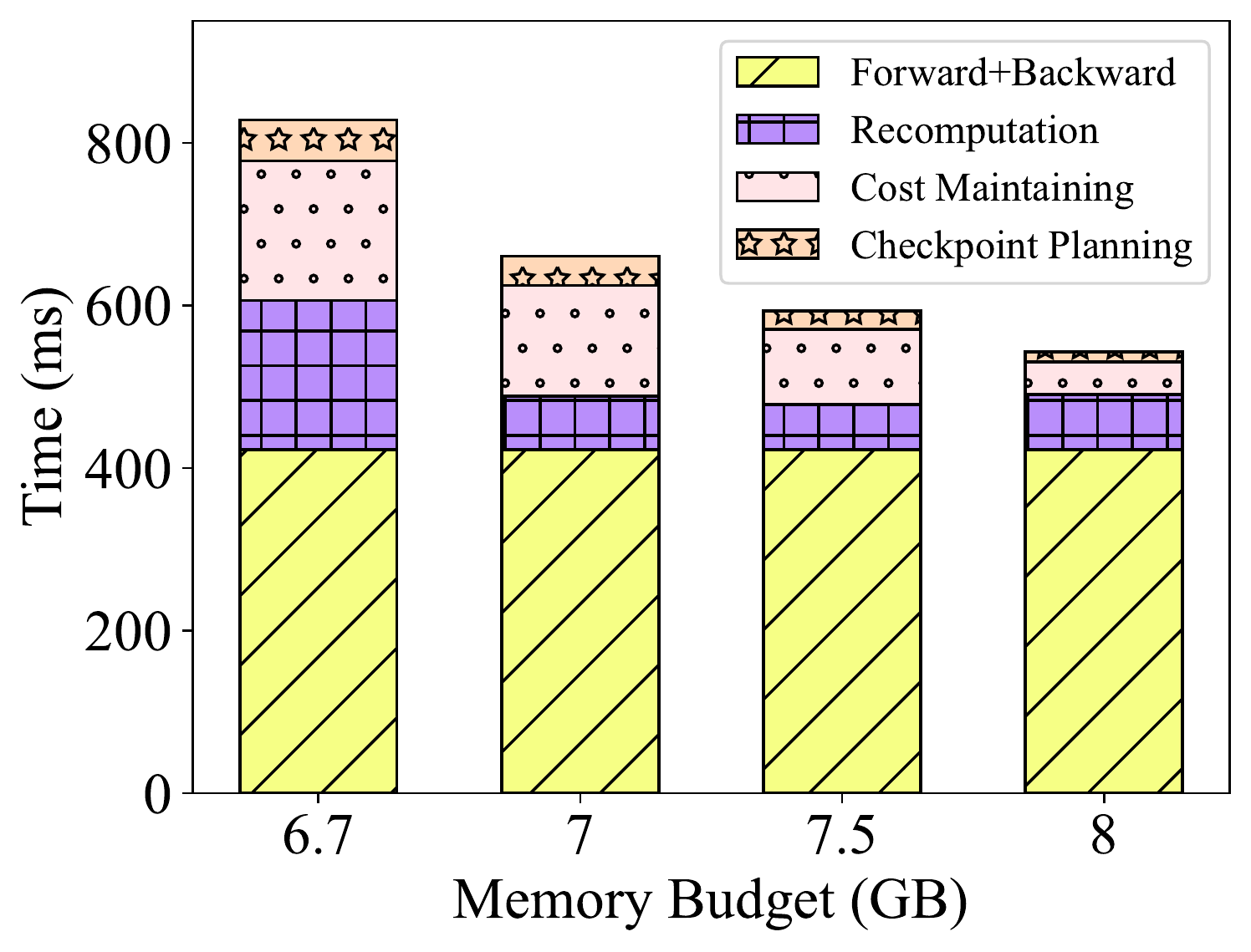}
		\caption{Training time breakdown of training Roberta-base on SWAG dataset using dynamic checkpointing planner, \textit{DTR}. The memory budget is set to 4.2/4.5/5/5.5 GB (actually 6.7/7/7.5/8 GB used).}
		\label{fig:dtr-overhead}
	\end{minipage} 
\end{figure*}

\subsection{Memory Impact of Dynamic Input Size}
\label{sec:motivation_1}

Input size is represented by the number of elements in the input tensor for each mini-batch.
The dynamic of input size comes from two aspects: dataset and data argumentation. 
Figure~\ref{fig:dataset-nlp} shows the input size distribution when training Bert-base on SWAG, SQuAD, and GLUE-QQP datasets (batch size of 16, 12, and 32, respectively).
Among different datasets, the range of input size is quite large, with 35$\sim$141, 153$\sim$512, and 30$\sim$332 for the above three datasets, respectively.
And the input size tends to follow a certain probability distribution, such as normal distribution and power-law distribution.

Consequently, the dynamic of input size can greatly affect the GPU memory footprint of activation tensors. 
The memory footprint during each training iteration consists of model parameters, gradients, optimizer states, and activation tensors, where the first three are constant regarding the model structure and do not change across different input sizes.
Figure~\ref{fig:dataset-nlp} also shows the GPU memory usage under various input sizes. With input size increasing, the memory usage increases accordingly. 
Besides, the GPU memory usage curve is quite smooth, revealing the possibility for accurate memory prediction with analytical models.

\subsection{Inefficiency of Current Checkpointing Planners}
\label{sec:motivation_2}
\textbf{Static checkpointing planners conservatively preserve memory for the largest input size and thus lead to low training throughput.} Training a large DL model successfully on GPU means no OOM exception happens throughout all training iterations. 
Thereby with the dynamic of input size, static checkpointing planners have to conservatively generate checkpointing plans regarding the peak memory usage of the largest input size.
Figure~\ref{fig:small-input-size-gap} illustrates the GPU memory usage of training Bert-base model on GLUE-QQP dataset (with batch size of 32) using \textit{Sublinear}~\cite{chen2016training} under the GPU memory budget of 3 GB. 
\textit{Sublinear} therefore generates the checkpointing plan targeting the maximum input tensor size (e.g., with \textit{seqlen}=300) to conservatively avoid OOM exception. Unfortunately, considering \textit{Sublinear} applies the above plan to a smaller input tensor (e.g., with \textit{seqlen}=55), it unnecessarily leaves 1.2 GB memory budget unused, which can be used for better training throughput. 
Even without checkpointing, the peak memory usage of a smaller input tensor may still be within the memory budget. Unnecessarily checkpointing in such cases sacrifices training throughput due to the conservation of static planners. As shown in Figure~\ref{fig:small-input-size-gap}, the reduced training throughput is non-trivial, which can be as large as 35\%. 

\textbf{Dynamic checkpointing planners redundantly generate plans for repeated input size, and thus lead to high checkpointing overhead. } Figure~\ref{fig:dtr-overhead} shows the training time breakdown of Bert-base on SWAG dataset using \textit{DTR}~\cite{dtr} under the GPU memory budget of 4.2/4.5/5/5.5 GB. However, \textit{DTR} uses 6.7/7/7.5/8 GB memory actually due to the severe memory fragmentation.
It is obvious that the planning (which activation tensor to be dropped/evicted) overhead takes 4.40\% of the overall iteration time on average. With a lower memory budget, this planning overhead can reach 6.06\% at most.
With a lower memory budget, the memory fragmentation becomes more significant, and thus leads to more evictions that tremendously increase the planning overhead.
As described in Section~\ref{sec:motivation_1}, the input size follows some probability distribution, and each input size can repeatedly appear during the training iterations. However, \textit{DTR} overlooks such characteristics of dynamic input sizes. It thus treats the input tensors independently, and redundantly generates the same checkpointing plans for the repeated input tensors, which leads to significant planning overhead.
Besides, the re-computation overhead in \textit{DTR} is also non-trivial, which can be up to 20.7\% of the training time.

\section{Design}
\label{sec:design}
\subsection{Design Overview}

\textit{Mimose} is designed to be dynamic and agile in generating and applying checkpointing plans according to the input size on the fly. Due to this design philosophy, \textit{Mimose} exists on the critical path of the training pipeline. Therefore, lightweight modules are designed for \textit{Mimose} to ensure low overhead.

\begin{figure*}[h]
	\centering
	\includegraphics[width=0.85\linewidth]{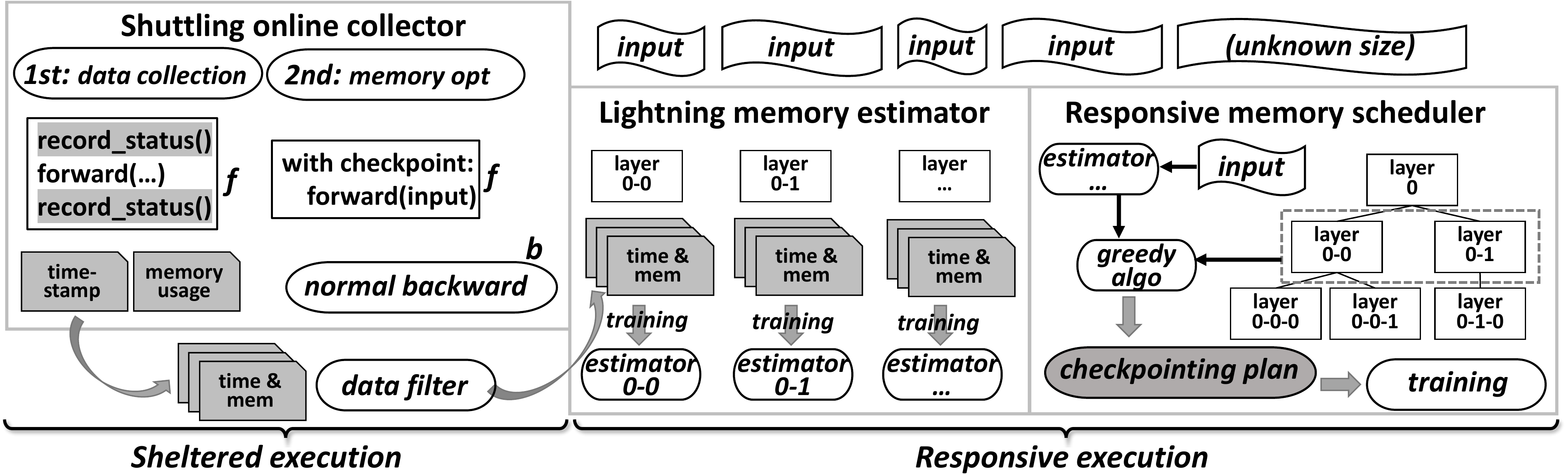}
	\caption{Design overview of \textit{Mimose}.}
	\label{fig:design-overview}
\end{figure*}

Specifically, \textit{Mimose} is mainly composed of a shuttling online collector, a lightning memory estimator, and a responsive memory scheduler, as shown in Figure~\ref{fig:design-overview}.
The \textbf{shuttling online collector} collects the memory usage and forward computation time of each layer in a given DL model (e.g., encoder, attention).  It can perform collection online without pre-analyzing the model even under insufficient GPU memory. 
The \textbf{lightning memory estimator} builds a memory prediction model based on the collected data and estimates the per-layer memory usage for each unknown input tensor size.
The \textbf{responsive memory scheduler} is responsible for exploring a near-optimal checkpointing plan based on the estimated memory consumption and the computation time and then scheduling the activation tensors with negligible overhead.

\textit{Mimose} divides the whole training process into two phases. 
During \textbf{1) sheltered execution}, \textit{Mimose} leverages the shuttling online collector, which considers the DL model as a sequence of building blocks (e.g., encoder block, attention block). It modifies the forward calculation per block and executes it twice in a training iteration. 
At the end of this phase, the memory consumption data is fed to the memory estimator to train the memory estimation model. 
In our experience, this phase requires only 10$\sim$30 iterations.
During \textbf{2) responsive execution}, \textit{Mimose} passes the argumented input tensor to the responsive memory scheduler. If there is a checkpointing plan of similar input size in its cache, the plan can be picked up directly. Otherwise, when cache miss occurs, the scheduler, together with the memory estimator, can derive the near-optimal checkpointing plan in less than a millisecond. In this phase, the online collector is frozen, and no additional knowledge is required.

\subsection{Shuttling online collector}

If with unlimited GPU memory, we can directly profile the model during the forward pass, for instance, compare the timestamp and memory footprint in different stages (e.g., before and after the forward computation of some layers) so that we can get the memory occupation and computation time of each layer.
However, given random-sized input tensors, the activation tensors can consume large memory and cause OOM exception. 
To ensure that the model can be trained properly, we need to apply the conservative checkpointing (e.g., \textit{Sublinear}) during memory/time data collection.
Besides, there is a contradiction between applying checkpointing and inspecting activation tensors: checkpointing means discarding the activation tensors instantly, and thus these tensors can be neither revisited nor inspected.

\begin{figure}[htbp]
	\centering
	\includegraphics[width=\linewidth]{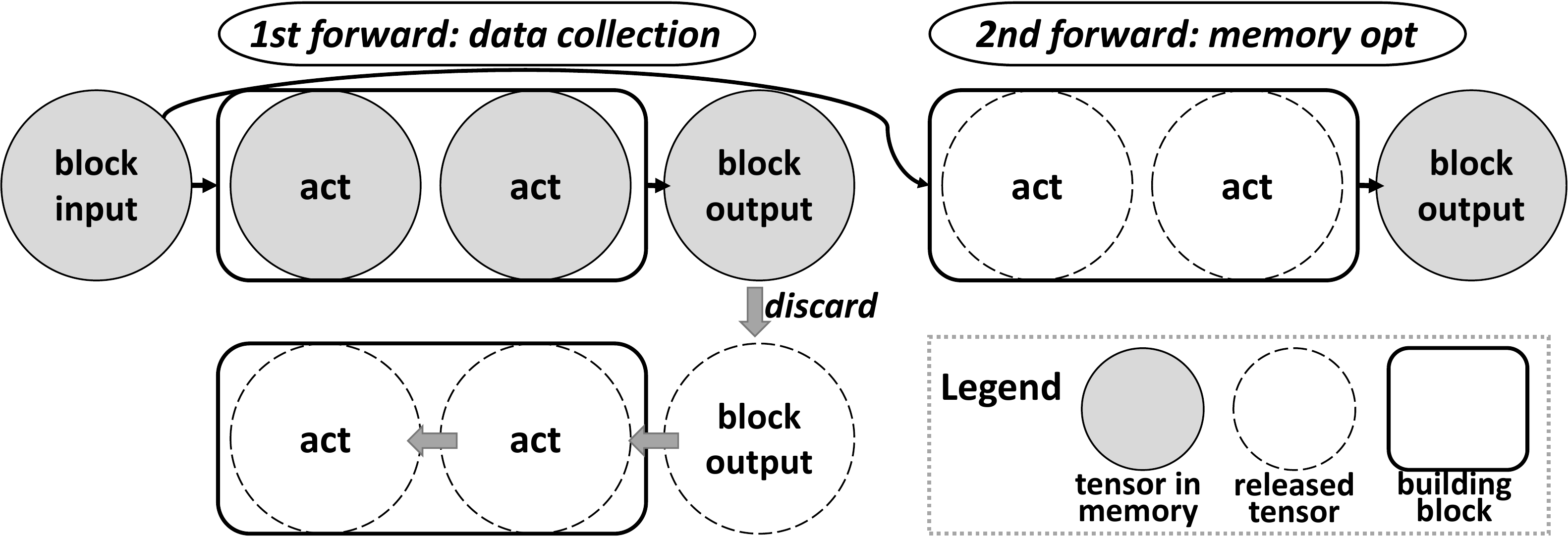}
	\caption{Two forward passes of the shuttling online collector.}
	\label{fig:forward-twice}
\end{figure}

Therefore, we propose the shuttling forwarding, where the forward pass of each layer will be executed twice, as in Figure~\ref{fig:forward-twice}. 
A DL model is split as a sequence of building blocks (e.g., encoder block, attention block). 
The first forward computation is conducted as normal, but the final output tensors inside this block are discarded, and the activation tensors are dropped consequently.
The second forward computation is conducted oppositely, with all activation tensors inside this block dropped instantly, except checkpointing the output tensor, so as to minimize the memory usage and prepare for the data collection for the next block. 
Note that the shuttling forwarding is conducted block by block, with the activation tensor between blocks kept in GPU memory, which also means that its memory footprint is the same as that of \textit{Sublinear} planner~\cite{chen2016training}. 

As for the time overhead, compared to \textit{Sublinear}, shuttling collector only repeats the forward pass in each training iteration. 
Additionally, compared to normal training without any memory planner, it takes more time for recomputation in each iteration. 
Since the time of holistic forward pass is generally shorter than that of backward pass, the overhead of shuttling collector is at most twice that of normal training iteration.
Although its overhead per iteration seems large, \textit{Mimose} requires a trivial number of iterations for collection. 
\textit{Mimose} uses shuttling collector only when meeting new input size, so the overhead can be reduced to $O(\frac{n}{N})$ throughout the training process, where $n$ represents the types of input size and $N$ represents the total iterations. 
The data provided by the shuttling collector is used to train the memory estimator model (Section~\ref{sec:memory-prediction}). 
If combined with a lightweight but accurate memory estimator model, the collector overhead can be reduced to a very low level.

\subsection{Lightning memory estimator}
\label{sec:memory-prediction}

The memory estimating model lies on the critical path of the training pipeline with \textit{Mimose}. Therefore, it should satisfy the following rules.
\begin{itemize}
\item It should require less training data, since its training data has to be collected online during the sheltered execution.
\item Its prediction should be fast enough, since its prediction is the prerequisite for generating checkpointing plan.
\item It should be accurate enough to provide memory usage information to the subsequent checkpointing scheduler for reasonable plans.
\end{itemize}

\begin{figure}[htbp]
	\centering
	\includegraphics[width=\linewidth]{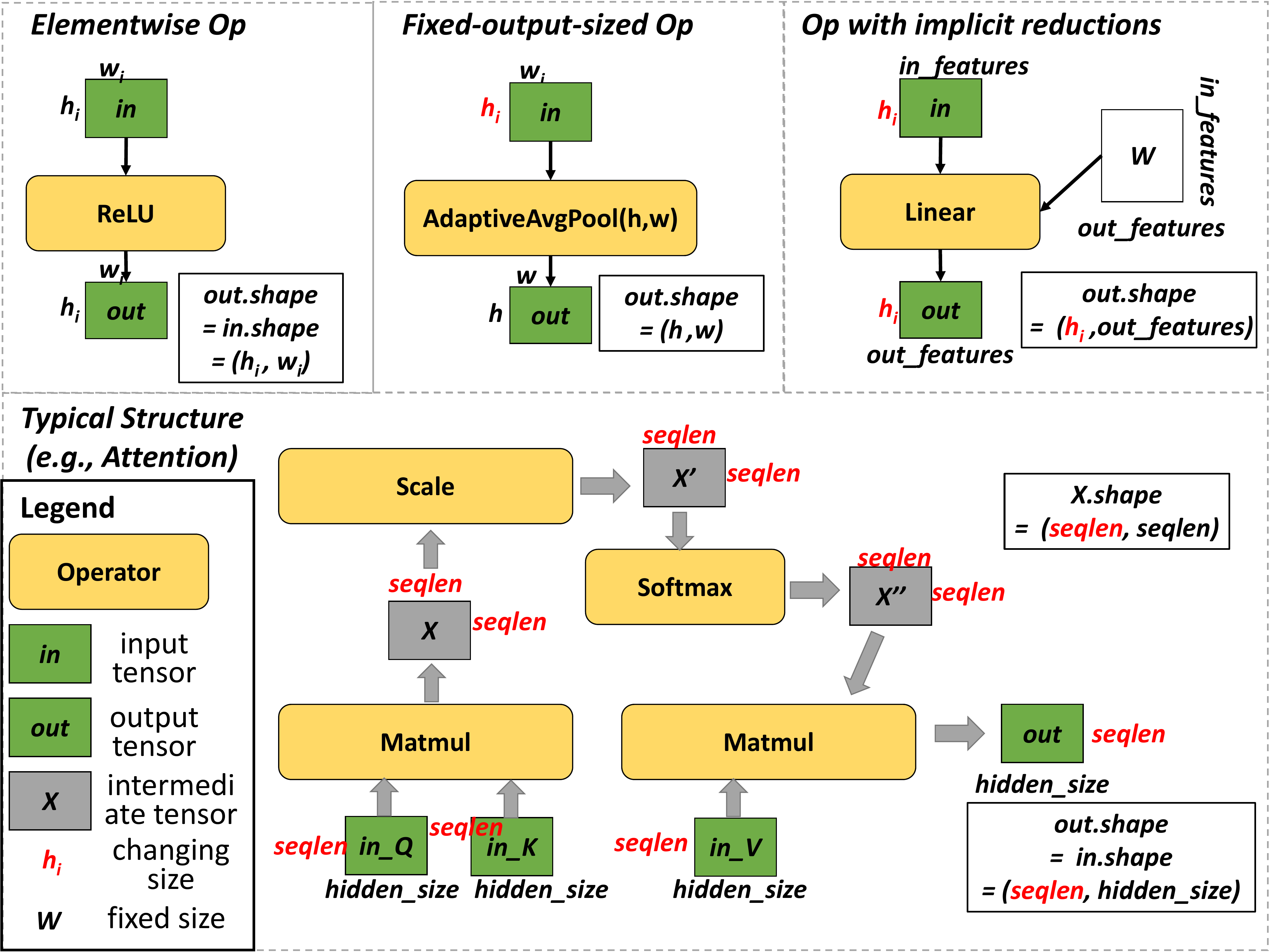}
	\caption{Relationship between input tensor shape and output tensor shape of four representative operators.}
	\label{fig:input-relationship}
\end{figure}

\begin{figure}[htbp]
	\centering
	\includegraphics[width=0.8\linewidth]{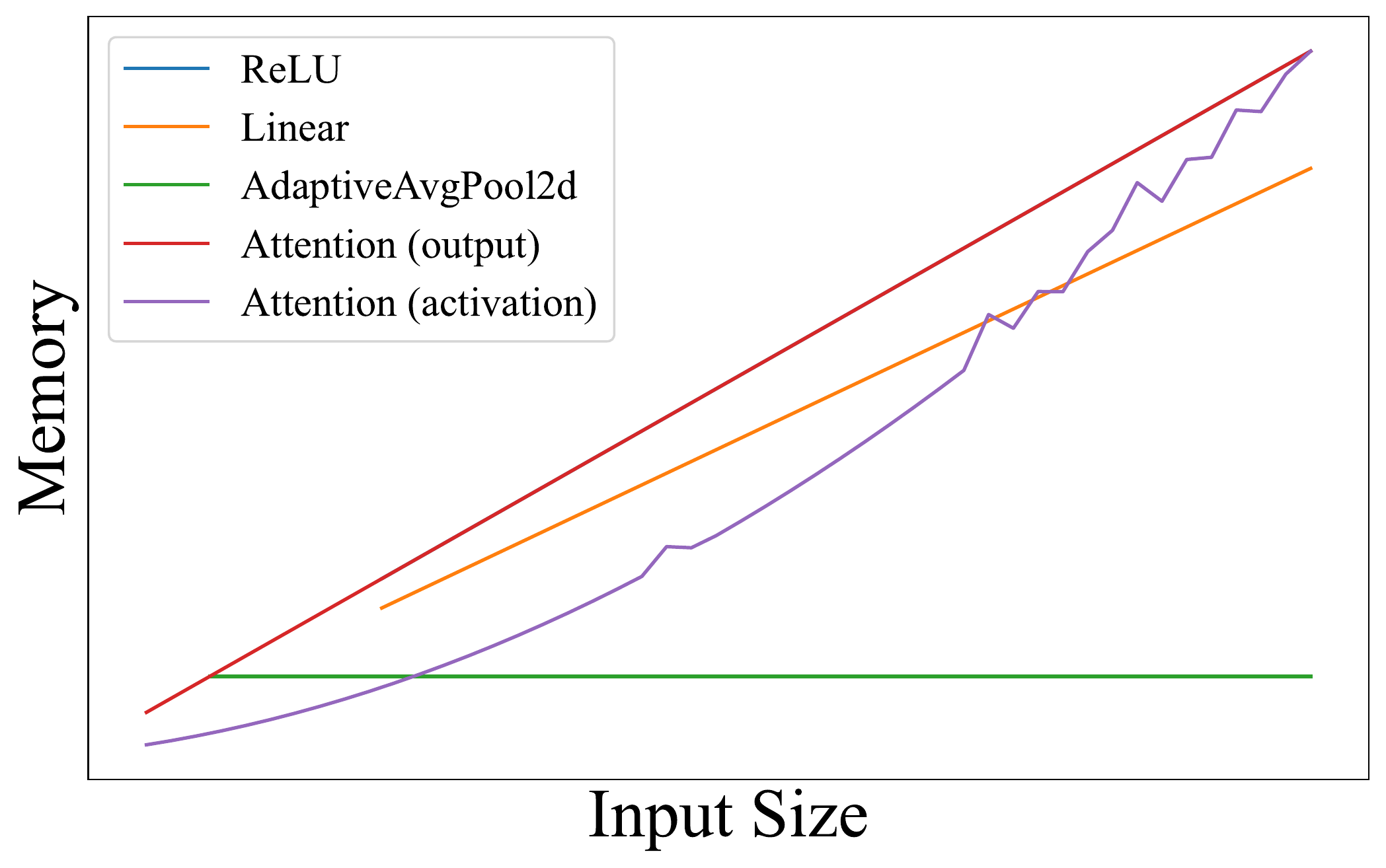}
	\caption{Correlations between output tensor size and input tensor size relationship of common layers/structures.}
	\label{fig:input-output-tensor}
\end{figure}

The activation tensors in model training are actually composed of the output tensors of all operators in the model. Given a static model, the number of tensors forming the activation is constant, so we should focus on the size of each tensor.
To construct the memory estimator model between input tensors and activation tensors, we study the relationship between them in representative DL operators, including both layers and neutral network structures, and we classify them into four categories, as shown in Figure~\ref{fig:input-relationship} and Figure~\ref{fig:input-output-tensor}. 

\textbf{Elementwise operators}, such as \texttt{ReLU} and \texttt{add}, perform individual operations on each element of the input tensor. Therefore, the output tensor shares the same size as the input. 

\textbf{Fixed-output-sized operators}, convert the input tensor to an output tensor with fixed size. 
For example, the \texttt{AdaptiveAvgPool} operator applies an adaptive average pooling over an input tensor and can output a tensor with a pre-defined size.

\textbf{Operators with implicit reductions}, which contain reduction operations as part of the operators, such as \texttt{Linear}, \texttt{GEMM}, \texttt{Convolution}, and \texttt{maxPool}. 
Specifically, as for \texttt{Linear}, \texttt{GEMM}, the iteration input tensor only determines their input tensor shape in only one dimension. While in other dimensions, their shapes are carefully designed by the DL experts after substantial hyper-parameter tuning, and thus are specially fixed during training.
Therefore, the input-output tensor shapes have a deterministically linear correlation. 
As for \texttt{Conv}, \texttt{maxPool} and other operators with shape changes in multiple dimensions, this relationship is slightly more complicated. 
But these dimension sizes have deterministic relationships with extra variables that are also specially fixed, such as stride, kernel size, and padding size,  
and thus the resulting output tensor size still has a linear correlation. 

\textbf{Typical structure with a set of operators} can bring more complex memory usage relationship, such as \texttt{attention}, as shown in Figure~\ref{fig:input-relationship}. 
The input $Q$, $K$, and $V$ has exactly the same shapes, i.e., $(seqlen, hidden\_size)$, where $hidden\_dims$ is specifically fixed in model, but $seqlen$ is linearly proportional to the iteration input size. 
After the first \texttt{Matmul} operation, $Q \times K^T$, a tensor with shape of $(seqlen, seqlen)$ is generated, which will increase the memory usage by $seqlen \times seqlen$. Later, after the \texttt{Scale} and \texttt{Softmax} operations, another two intermediate tensors with size of $(seqlen, seqlen)$ are generated.
Therefore, the sizes of these intermediate tensors are quadratically correlating to the iteration input size.
And one of them is multiplied with input tensor $Q$ whose shape is $(seqlen, hidden\_size)$, to get a tensor with shape of $(seqlen, hidden\_size)$ as the final output. It is obvious that the output tensor size also has a proportional relationship with the input tensor, $Q$, $K$, or $V$. 
For this reason, the input tensor of other subsequent layers is still linearly correlated to the iteration input tensor, therefore, avoiding the size explosion due to function compositions.

From the above analysis, we find that the sizes of activation tensors in a general DL model are almost polynomially correlated to the size of input tensors, and it is at most quadratic in most cases. 
Moreover, the input size of each operator (both individual operators and structures) should be linearly correlated to the input tensor size of the mini-batch (i.e., current iteration) so that the memory usage of activation tensors can be abstracted a lightweight polynomial function.

Based on the above study, we finally choose the polynomial regression model in the memory estimator. Compared with other complex algorithms, such as XGBoost~\cite{xgboost}, it achieves a satisfying trade-off between accuracy and efficiency, and we will evaluate a variety of fitting algorithms in Section~\ref{sec:exp-prediction-model}.

Although the polynomial model works well for NLP models, there are still some structures that do not conform to polynomial correlation especially for object detection models. For example, the \texttt{padding} operation in Swin-Transformer, which targeted to solve the mismatching between input size and window size, has a step effect on the output tensor size. In general, this operation can cause a fluctuating error of no more than 5\%, whose impact is not significant luckily.
However, another model design, i.e., the 2-stage object detection, may have a greater impact. The number of generated anchors/proposals is not fixed, because they are mainly related to the content (e.g., how many people in an image sample) of the input tensor. 
Since this fluctuation is unpredictable, it could have a negative impact on the memory estimator. Therefore, we leave the support of objective detection models for our future work, and we plan to apply some adaptive algorithms to the memory estimator.

\subsection{Responsive memory scheduler}

\begin{figure}
	\centering
	\subfloat[On Swin-Transformer blocks]{
		{\includegraphics[width=0.42\linewidth]{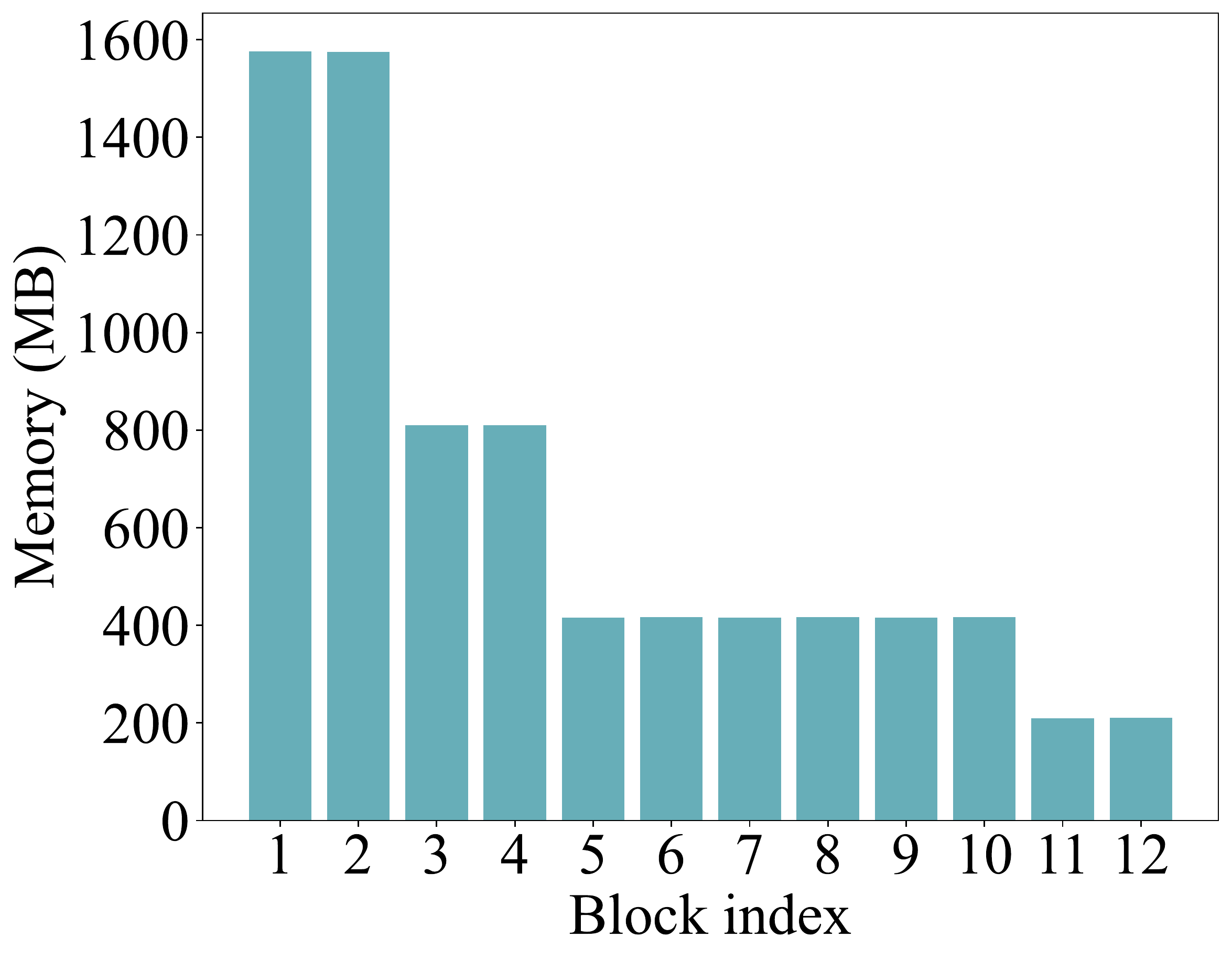} \label{fig:swin-block-memory}}
	}
	\subfloat[On ResNet BottleNecks]{
		{\includegraphics[width=0.42\linewidth]{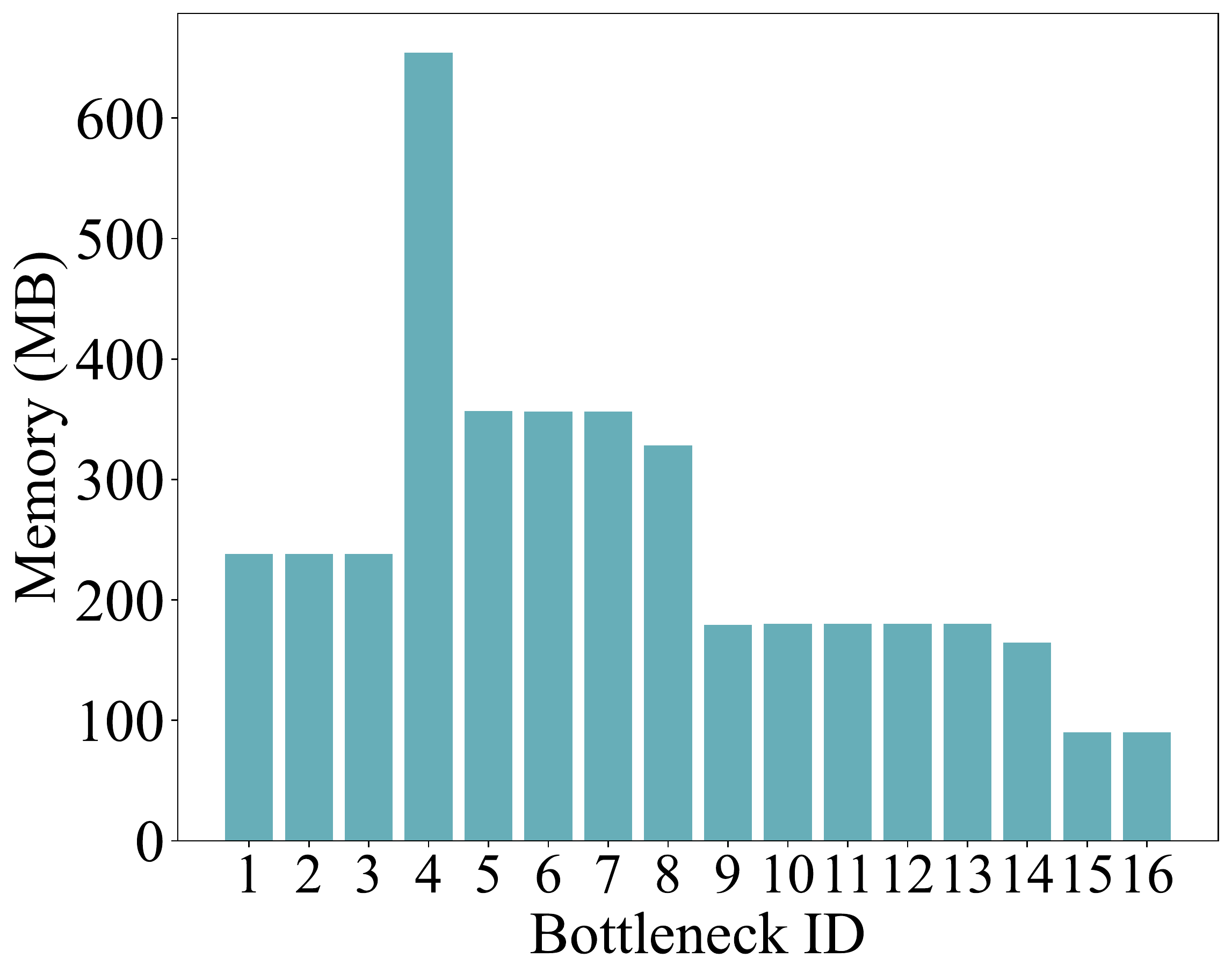} \label{fig:resnet-bottleneck-memory}}
	}
	\caption{Memory usage of activation tensors.}
\end{figure}

The same structures in a model could have different memory usages. 
For simplicity, we define the \textit{stage}, which represents a set of layers corresponding to the user-written model code structures, and we regard \textit{stages} as natural separators to locate layers in the model.
Take Swin-Transformer for example, as shown in Figure~\ref{fig:swin-block-memory}, the patch merging structure on the boundary of each stage reduces the output tensor size of the previous stage by 50\%.
This structure leads to the step-down of memory usage in different stages.
However, the first stage of ResNet has a different structure from the other stages, which does not show the same step-down trend, as shown in Figure ~\ref{fig:resnet-bottleneck-memory}.

\begin{figure}[htbp]
		\centering
		\includegraphics[width=0.9\linewidth]{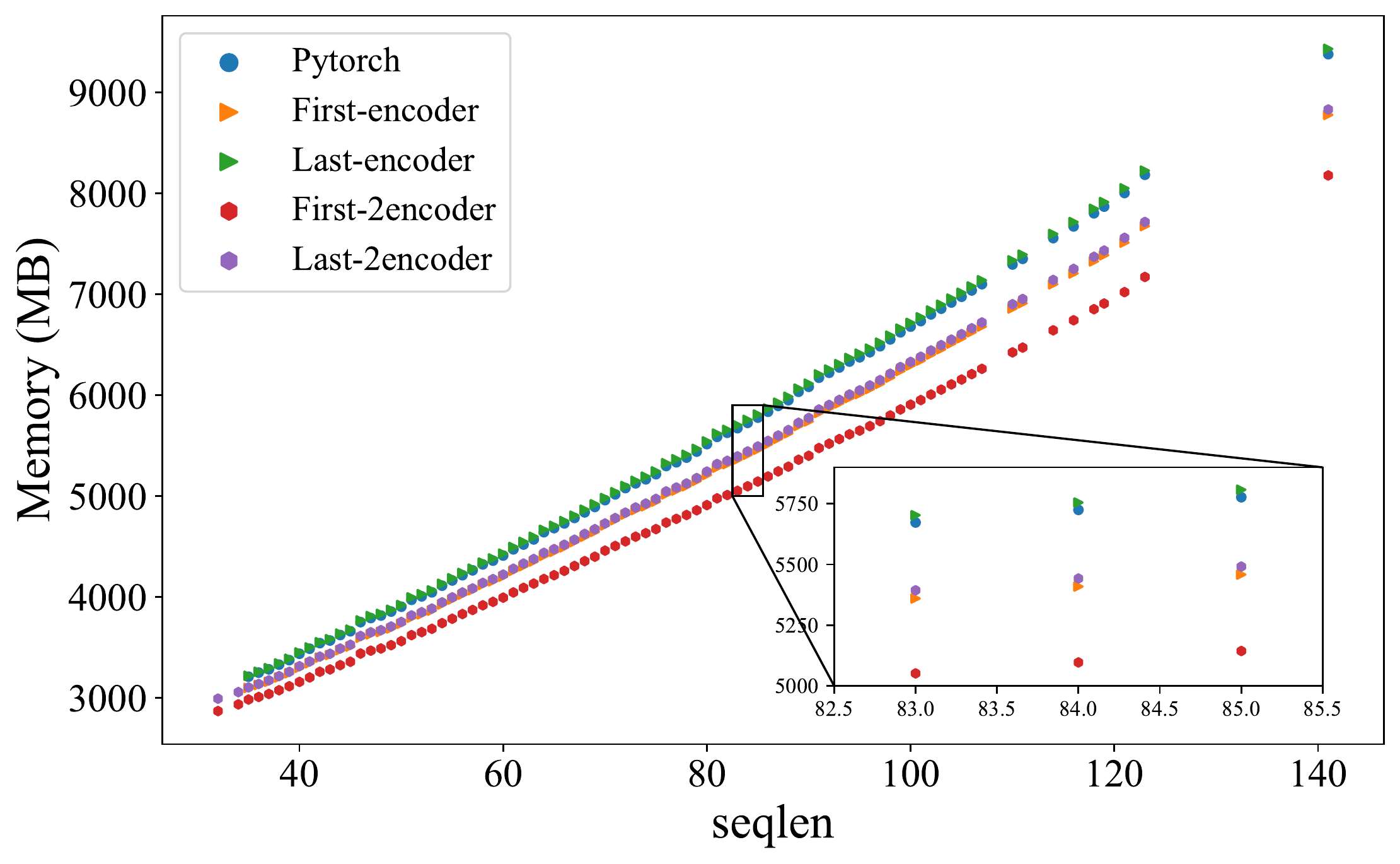}
		\caption{Peak memory usages of checkpointing different encoders in Bert-base, which mainly contains 12 encoders in total.}
		\label{fig:checkpoint-diff-layer}
\end{figure}

In addition, even for structures with the same memory usage, checkpointing at different structures can still lead to different peak memory usages. 
Bert-base mainly contains 12 encoders, whose activation tensors are the main contributing part to memory usage, and each encoder consumes the same amount of memory. 
When we apply checkpointing to one encoder, no matter which encoder is selected, the peak memory usage and the memory usage at the end of the forward pass keep constant. 
In backward pass, we have to perform recomputation to restore the encoder's activation tensors. However, different encoder selections could result in different peak memory usages. 
Figure~\ref{fig:checkpoint-diff-layer} shows the peak memory usages of checkpointing different encoder structures under various input tensor sizes (i.e., \textit{seqlen}). 
Due to the characteristic of backward computation, the firstly forwarded encoder is backwarded lastly, and the lastly forwarded encoder is backwarded firstly.
If the checkpointed encoder is the last encoder in the model, it means that this encoder's activation has to be restored instantly once the backward computation starts. At this time, activation tensors of other encoders are not released in order to participate in the subsequent backward computation.
It leads to a high peak memory usage, which is similar to the scenario without any checkpointing at all. 
If the checkpointed encoder is at the front of the model, it means that this encoder's activation has to be restored instantly at the end of the backward pass. At this time, activation tensors of most encoders are already released, which has a trivial impact on peak memory usage.
Therefore, we prefer checkpoint layers/structures with earlier timestamps in forward pass when their activation tensors have similar sizes.

Based on the above observations, we adopt a greedy algorithm for checkpointing scheduling, as shown in Algorithm~\ref{alg:greedy}. 
We first derive the estimated memory usage of the given input tensor, leveraging the lightning memory estimator (line~\ref{line:estimator}).
And we assign the layers with similar estimated memory usage ($\pm 10\%$ in our implementation) to a bucket and sort them according to the execution sequence in forward pass (line~\ref{line:bucket_start}$\sim$\ref{line:bucket_stop}). 
Then we select the layers that need to be recomputed one by one according to their activation sizes (line~\ref{line:select_start}$\sim$\ref{line:select_stop}). 
When the excess memory cannot be covered by one layer, the remaining layer with the largest activation is selected as soon as possible (line~\ref{line:select_1}).
Otherwise, the layer whose activation size is nearest to the excess memory is selected (line~\ref{line:select_2}). 
Note that we prefer to select layers with earlier timestamps within a bucket in order to further reduce the peak memory footprint.
The selection procedure loops until the activation size of selected layers exceeds the targeted excess memory.


\begin{algorithm} 
	\caption{Greedy scheduling of the responsive scheduler.}
	\label{alg:greedy}
	\renewcommand{\algorithmicrequire}{\textbf{Input:}}
	\renewcommand{\algorithmicensure}{\textbf{Output:}}
	\begin{algorithmic}[1]
		\REQUIRE Memory budget $M$, input tensor size $x$, layer set $L$
		\ENSURE Set of dropped/recomputed layers $L'$ 
		\STATE est\_mem $\gets$ MemoryEstimator($x$) \label{line:estimator}
		\STATE buckets $\gets$ empty list \quad// Buckets of layers
		\STATE sorted($L$, key=<activation tensor size>, order=desc)
		\WHILE {! $L$.is\_empty()} \label{line:bucket_start}
			\STATE $l$ $\gets$ $L$.top()
			\STATE $L$.remove($l$)
			\STATE initialize new bucket with $l$
			\WHILE {est\_mem[$L$.top()] > est\_mem[$l$] $\times$ (1 - 10\%)}
				\STATE $l'$ $\gets$ $L$.top()
				\STATE bucket.append($l'$) \textbf{and} $L$.remove($l'$)
			\ENDWHILE
			\STATE sorted(bucket, key=<forward timestamp, order=asc)
			\STATE buckets.append(bucket)
		\ENDWHILE \label{line:bucket_stop}
		\STATE excess\_mem $\gets$ ($\sum${est\_mem} - $M$) \label{line:select_start}
		\WHILE {excess\_mem > 0}
			\STATE bucket\_candidates $\gets$ $\forall b \in \mbox{buckets}$, s.t.  max(est\_mem[$b$]) > excess\_mem
			\IF {bucket\_candidates.empty()}
				\STATE $l \gets$ buckets.top().top() // layer with largest activation\label{line:select_1}
			\ELSE
				\STATE $l$ $\gets$ bucket\_candidates.top().top() \label{line:select_2}
			\ENDIF
			\STATE $L'$.append($l$) \textbf{and} remove $l$ from buckets
			\STATE excess\_mem $\gets$ excess\_mem - est\_mem[$l$]
		\ENDWHILE \label{line:select_stop}
	\end{algorithmic} 
\end{algorithm}

\section{Implementation}
\label{sec:implementation}

The implementation of \textit{Mimose} is on the basis of the checkpointing API (i.e., \texttt{torch.utils.checkpoint}) provided by PyTorch (since v0.4.0), so that \textit{Mimose} is compatible with broad range of training codes written with PyTorch.
Although this also means being difficult to achieve tensor-level memory planning, it brings considerable performance benefits to \textit{Mimose} in generating and applying ever-changing checkpointing plans.
It is essential in scenarios with input tensor size dynamics, because \textit{Mimose} is lying on the critical path of the training pipeline. 

\begin{figure}[htbp]
	\centering
	\includegraphics[width=0.8\linewidth]{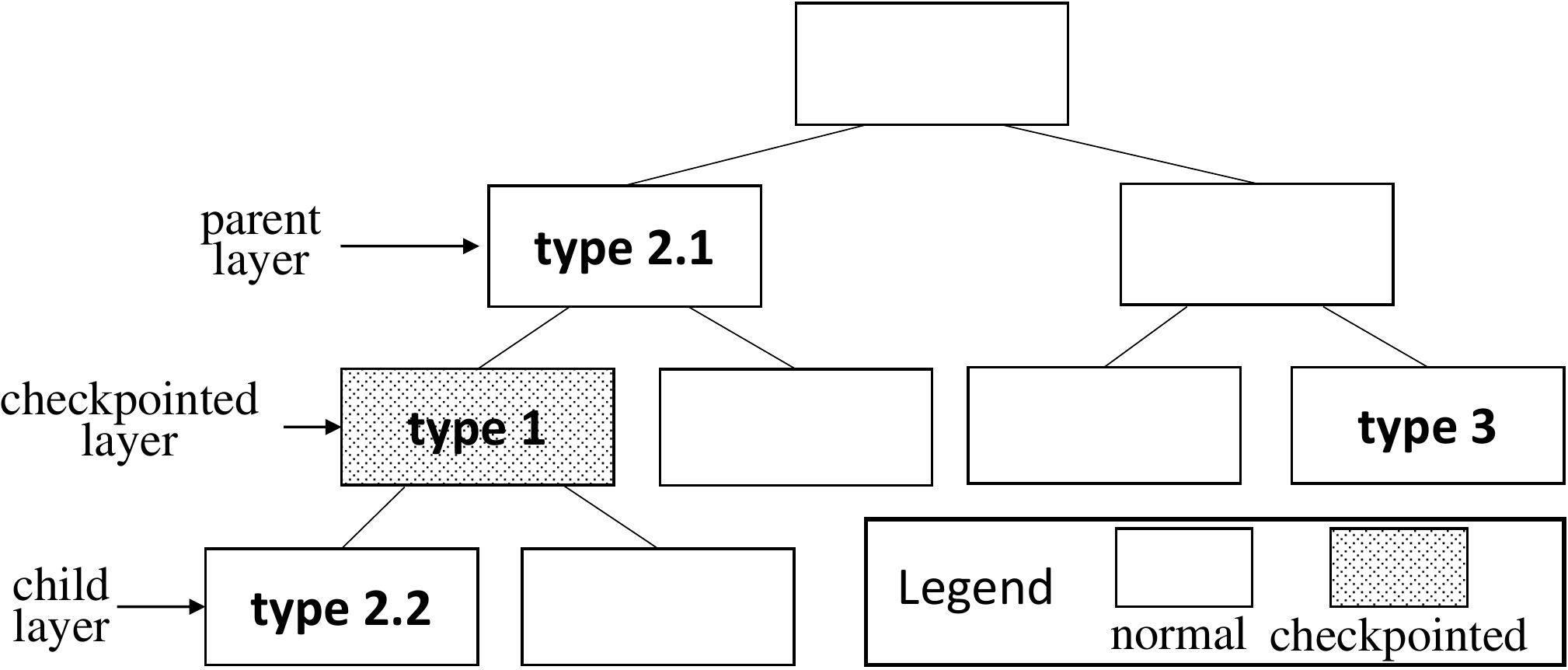}
	\caption{Illustration of data filter in shuttling online collector.}
	\label{fig:data-filter}
\end{figure}

In sheltered execution, the data collector needs to collect per-layer memory usage and per-layer forward computation time during the model training online. 
And it wraps the forward pass and instruments before and after the forward computation of each layer. Therefore the memory usage and computation time can be derived by comparing the state differences. 
During collection, the shuttling forwarding mechanism executes the forward pass of the building blocks (a DL model is split as a sequence of building blocks) twice, one for data collection, the other for minimizing memory usage by checkpointing the output tensors and dropping activation tensors.
However, due to the eager mode of PyTorch, there is no computation graph provided. Therefore, during the forward computation of a layer, the data collector cannot distinguish whether its parent layer or the child layers is checkpointed (i.e., under \texttt{torch.no\_grad()} context), and thus the collected data contains invalid items. 
Therefore, as shown in Figure~\ref{fig:data-filter}, a data filter is designed for three typical cases.
\textit{1)} If current layer is checkpointed, the data should be removed, since no activation tensor exists.
\textit{2)} If current layer is not checkpointed, but its parent layers or child layers are checkpointed, the data should be removed.
\textit{3)} Otherwise, the data is valid.

In responsive execution, the memory scheduler holds a cache to store the generated checkpointing plan and uses the input tensor size as the indexing key. 
Whenever an input size is encountered, the scheduler firstly searches in cache, so as to avoid the overhead of generating plans repeatedly.
In addition, the memory usages of similar input sizes are similar, and the generated plans are also similar. Therefore, they can also be the plans of each other.
During the forward pass, \textit{Mimose} looks for whether the ID of current layer exists in the previously generated checkpointing plan and determines whether to checkpoint accordingly, whose overhead is negligible. Compared with other checkpointing planners, this implementation enables fast and agile switching of plans.

\section{Evaluation}
\label{sec:evaluation}
\subsection{Evaluation Setup}

\noindent\textit{\textbf{{Hardware and software configurations.}}}
We conduct the experiments on a platform equipped with two-socket Intel Xeon E5-2680v4 CPUs (28 cores in total) and two NVIDIA V100 GPUs. 
The software environment contains Ubuntu 20.04 LTS operating system, CUDA toolkit 11.3, cuDNN v8.2.0, PyTorch v1.11, and HuggingFace transformers v4.18.0.

\noindent\textit{\textbf{{Tasks, datasets, and models.}}}
We evaluate \textit{Mimose} on four state-of-the-art NLP tasks, and their datasets, models, and batch size settings are shown in Table~\ref{table:model}. 
Specially, we use the representative datasets, SWAG~\cite{zellers2018swag}, SQuAD~\cite{rajpurkar2016squad}, and GLUE-QQP~\cite{wang2019glue} for the multiple choice, question answering, and text classification tasks, respectively.
And the NLP models involves Roberta-Base~\cite{liu2019roberta}, XLNet~\cite{yang2019xlnet}, and Bert-Base~\cite{devlin2018bert}, which contains 125, 110, and 110 million parameters. 

\noindent\textit{\textbf{{Comparison methods.}}}
We compare \textit{Mimose} with the static checkpointing planner \textit{Sublinear}~\cite{chen2016training} and the dynamic checkpointing planner \textit{DTR}~\cite{dtr} under various GPU memory budgets.
And we adopt the original PyTorch without checkpointing as the \textit{baseline}.  
The original \textit{DTR} implementation lacks support for several critical operators and fails to execute the above tasks. 
Therefore, we have extended \textit{DTR} with a few operators, such as \texttt{cumsum}, \texttt{split}, \texttt{masked\_fill}, etc.
However, since the implementation of operators such as \texttt{index\_select} used in XLNet is too special to be supported by \textit{DTR} for the time being, we fail to execute the QA-XLNet task with \textit{DTR}. 
We have also tried to compare with \textit{Checkmate}~\cite{checkmate}. However, it is implemented based on static graphs, and there are many problems in converting the models of the above tasks to static graphs. For example, the coverted static graph fails to tackle the input tensor with dynamic size.
Besides, \textit{Checkmate} is built on TensorFlow and thus cannot be compared with PyTorch-based implementations. For above reason, we omit \textit{Checkmate} for comparison.

\begin{table}[htbp]
	\centering
	\footnotesize
	\caption{Training tasks for evaluation.}
	\label{table:model}
	\begin{tabular}{ c|c|c|c }
		\hline
		Task & Dataset & Model & Batch Size \\
		\hline
		\tabincell{c}{Multiple Choice \\(MC-Roberta)} & SWAG & Roberta-B & 16 \\ \hline
		\tabincell{c}{Question Answering \\(QA-XLNet)} & SQuAD & XLNet & 16 \\ \hline
		\tabincell{c}{Question Answering \\(QA-Bert)} & SQuAD & Bert-B & 12 \\ \hline
		\tabincell{c}{Text classification \\(TC-Bert)} & GLUE-QQP & Bert-B & 32 \\
		\hline
	\end{tabular}
\end{table}

\subsection{Overall Performance}

\begin{figure}
	\centering
	\subfloat[Multiple Choice]{
		{\includegraphics[width=0.49\linewidth]{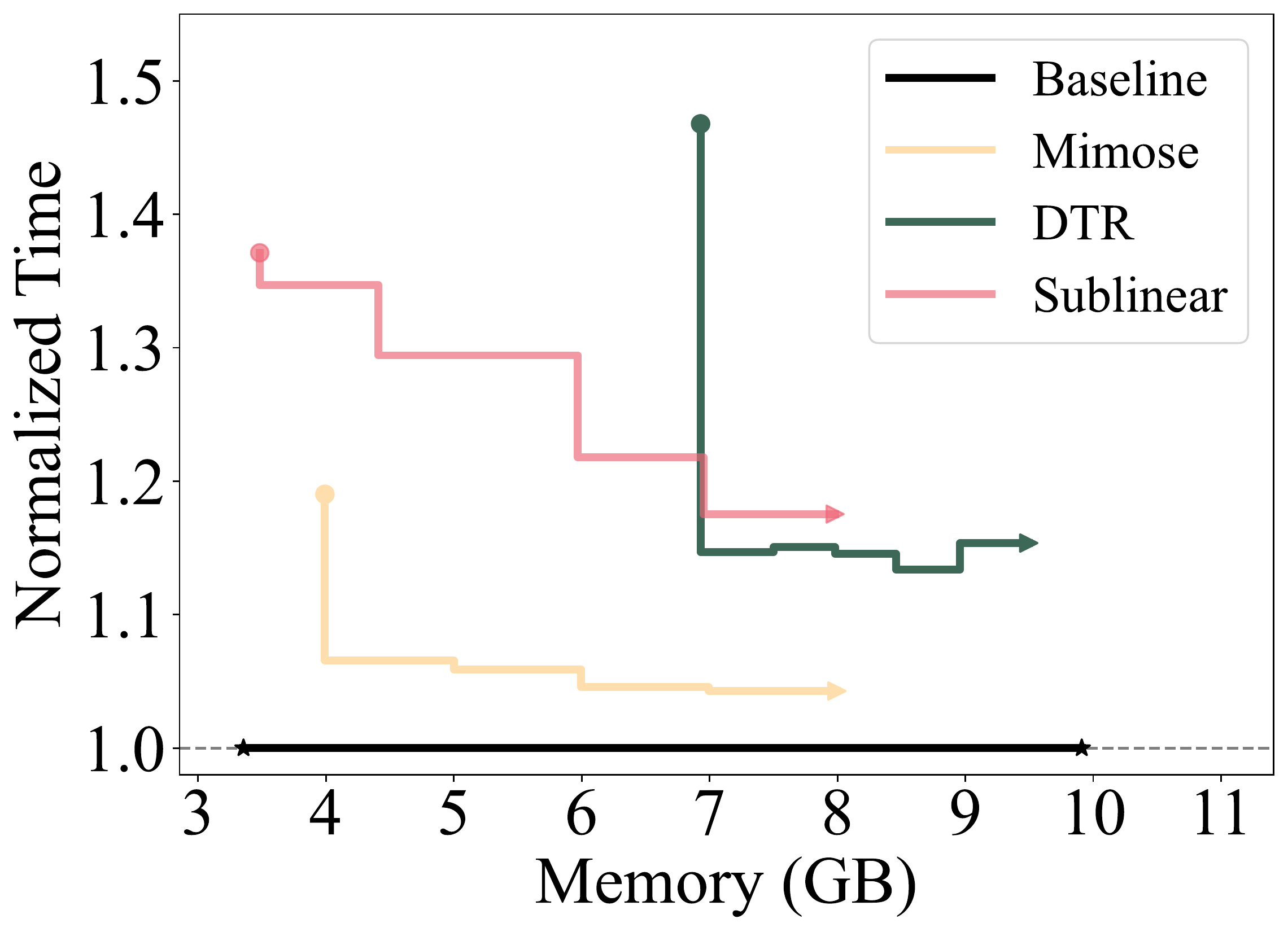} }
	}
	\subfloat[Question Answering (XLNet)]{
		{\includegraphics[width=0.49\linewidth]{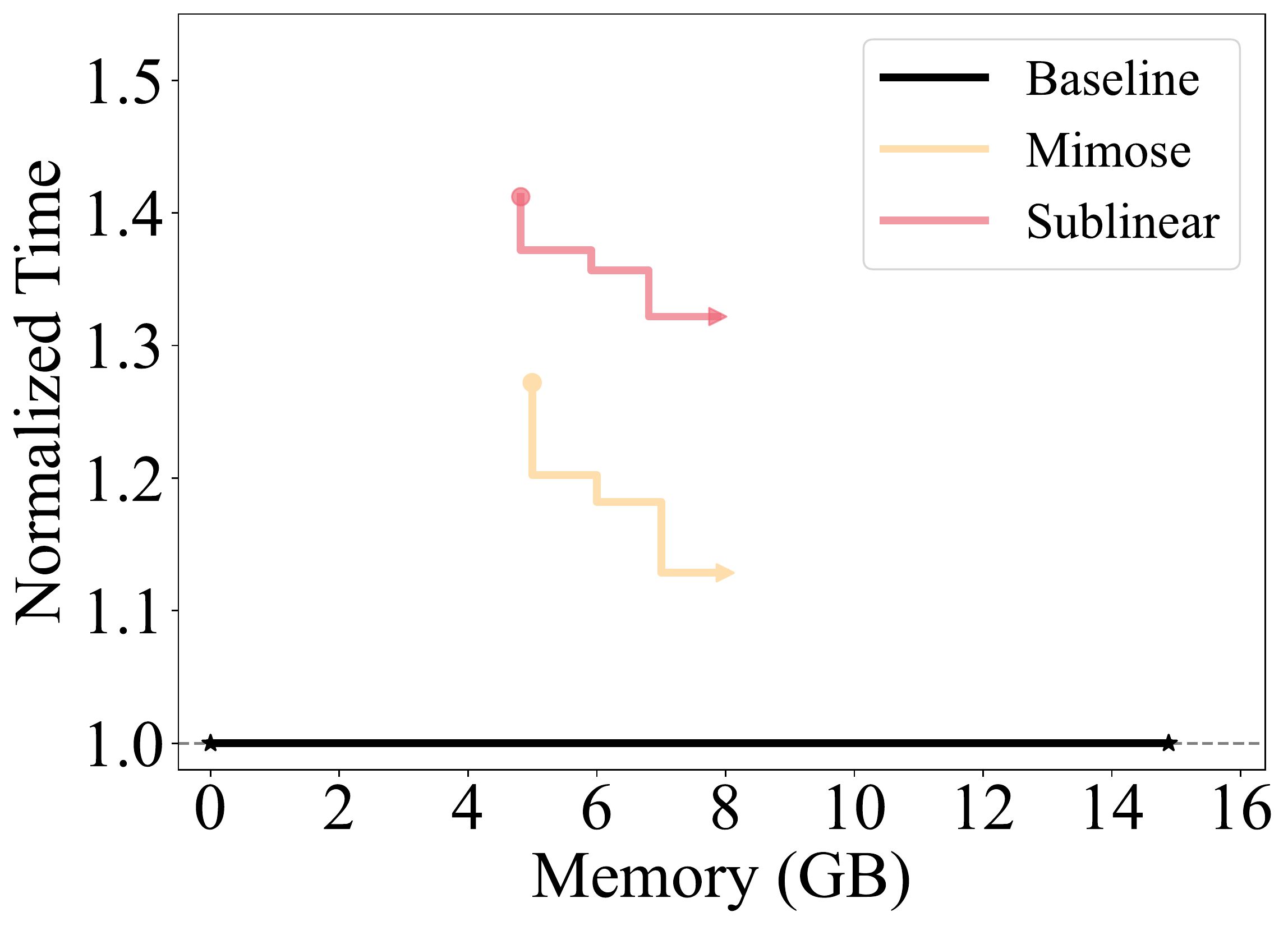} }
	}
	\vspace{-0.1in}
	\subfloat[Question Answering (Bert)]{
		{\includegraphics[width=0.49\linewidth]{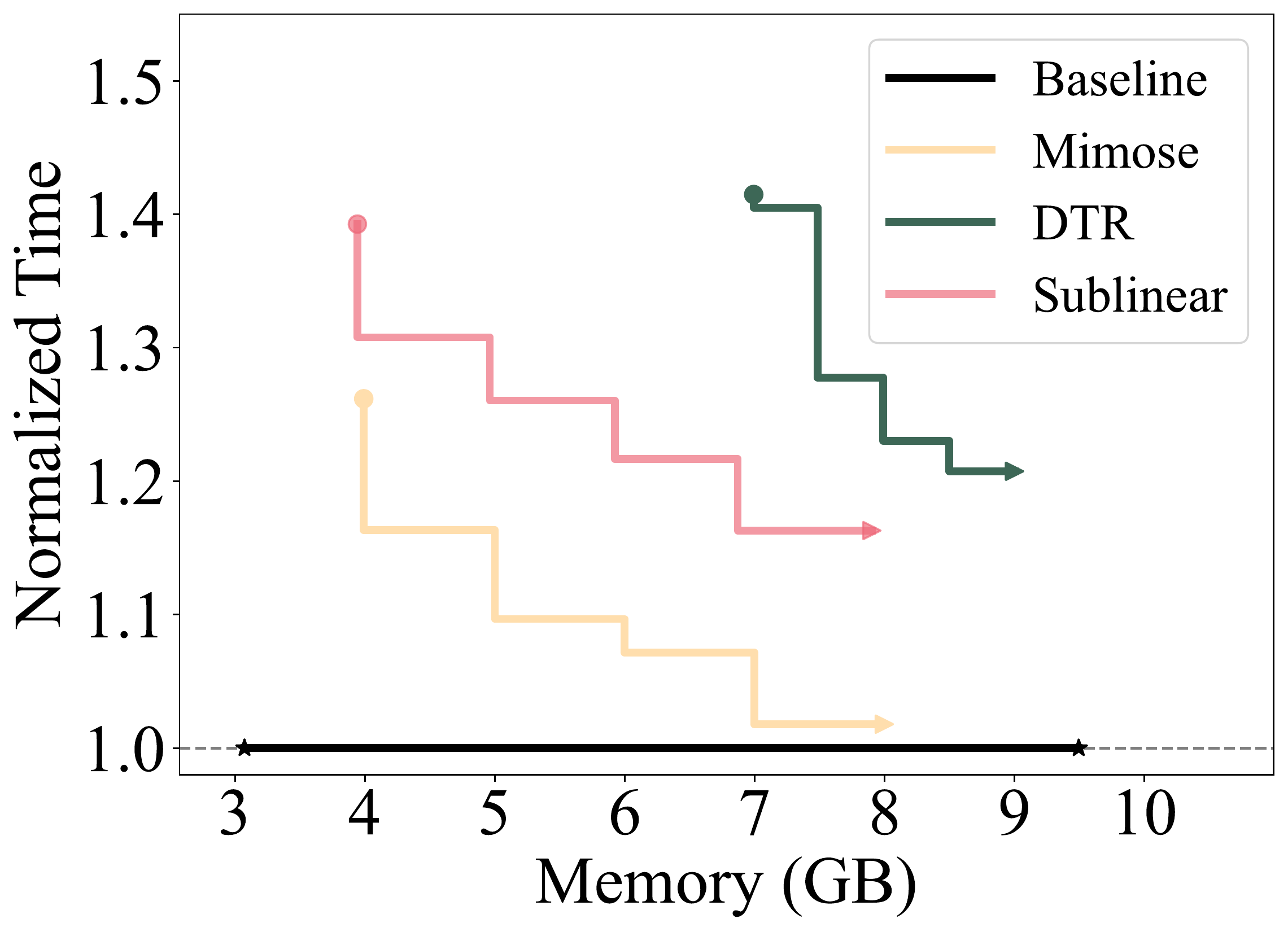} }
	}
	\subfloat[Text Classfication]{
		{\includegraphics[width=0.49\linewidth]{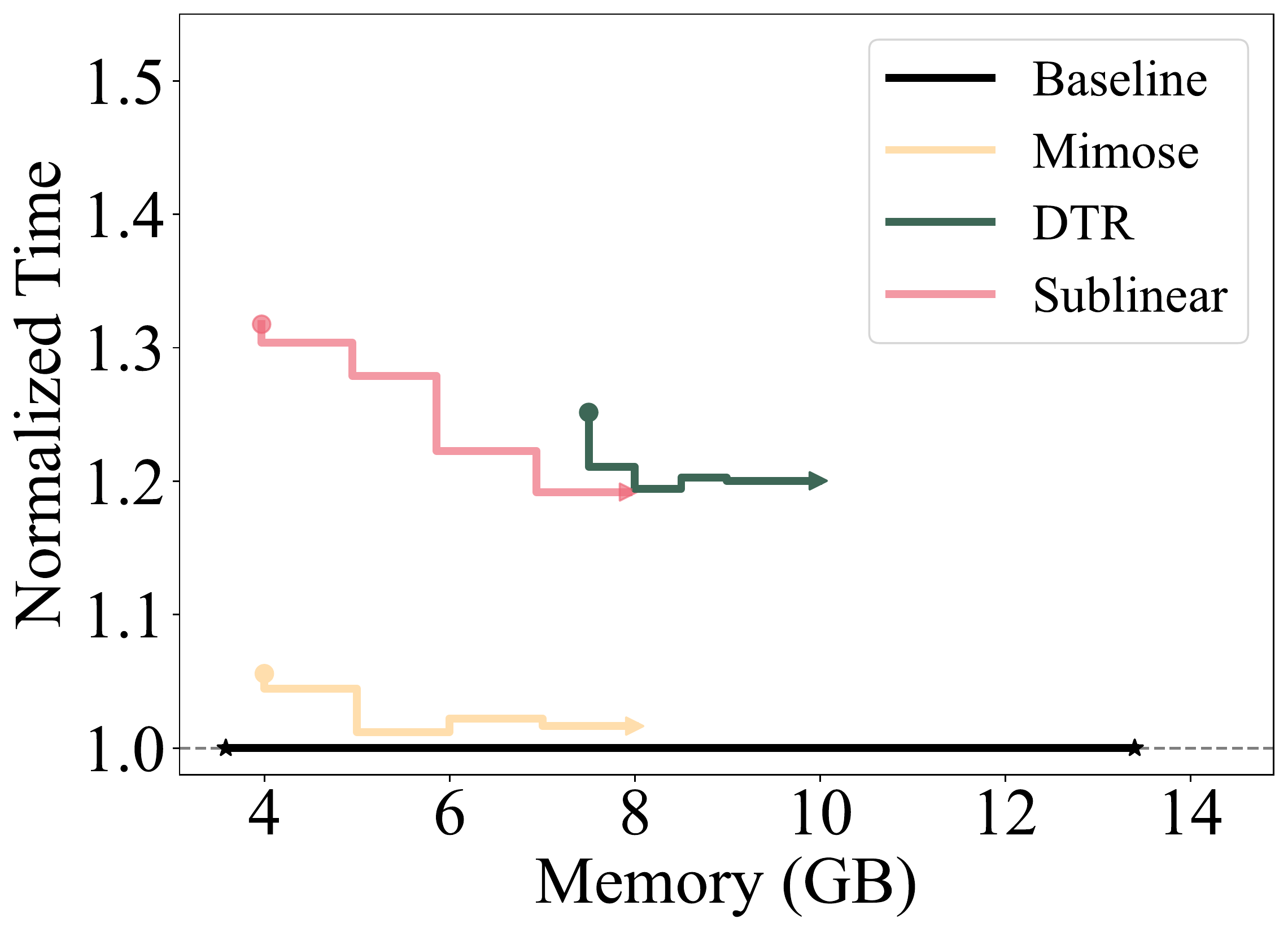} }
	}
	\caption{Single-epoch times for different methods normalized to \textit{Baseline} (original PyTorch without memory limit), where x-axis represents the memory budget. The ``star'' marks indicate the upper limit (all with checkpointing) and the lower limit (all without checkpointing) of the memory allocation size.}%
	\label{fig:overhead-overview}%
	
\end{figure}

To comprehensively evaluate different planners, we present the execution times under different memory budgets. Figure~\ref{fig:overhead-overview} shows the single-epoch times for different planners normalized to \textit{Baseline} (original PyTorch without memory limit). The ``star'' marks indicate the upper limit (all with checkpointing) and the lower limit (all without checkpointing) of the memory allocation size. As seen, \textit{Mimose} significantly outperforms \textit{Sublinear} with about 17.1\% improvement. This is because \textit{Sublinear} can only generate a static checkpointing plan based on the largest input tensor to avoid OOM, causing amounts of redundant recomputations for small input tensors. In contrast, \textit{Mimose} can adaptively generate plans according to input tensors to minimize performance overhead.

Compared with the dynamic checkpoint planner \textit{DTR}, \textit{Mimose} improves the performance by about 15.0\% on average. The reasons can be attributed to the following points. Firstly, the checkpointing delay of \textit{DTR} accounts for a high proportion of iteration time due to on-demand triggering by OOM. In addition, \textit{DTR} incurs a large search cost by repeatedly generating checkpoint plans for the same input sizes. Finally, \textit{DTR} generates lots of memory fragmentation at runtime, thus generating sub-optimal checkpointing plans with redundant computations. For the \textit{MC-Roberta} task, \textit{DTR}'s memory fragmentation reaches 2.5 GB with a 7 GB memory budget, whereas \textit{Mimose}'s is only 0.5 GB. A similar conclusion has also been proved in~\cite{hu2022megtaichi}.

It can be observed that the performance of \textit{Mimose} improves as the memory budget increases. For example, \textit{Mimose} achieves only 5.1\% slowdown compared to \textit{Baseline} under the memory budget of 8 GB. Furthermore, Mimose can still guarantee normal execution under a memory budget close to the lower bound (e.g., 3.36 GB for the \textit{MC-Roberta} task). The above results indicate that \textit{Mimose} can adjust the checkpointing plan for various input tensors, and reuse the plan generated for previous input sizes to reduce the data collection overhead.

\subsection{Overhead Breakdown}

We normalize the \textit{Mimose} overhead to the duration of the training task executing one iteration. Table~\ref{table:overhead-breakdown} shows the overhead breakdown under 6 GB memory budget when executing one epoch, where the total overhead is normalized to the single-iteration time. The \textit{Mimose} overhead comes from three parts: data collector, memory estimator, and memory scheduler. Specifically, the data collector involves redundant computations caused by executing each forward layer twice. The memory estimator predicts the per-layer memory usage for each input tensor size. The memory scheduler generates a checkpointing plan based on the estimated memory usage and schedules the activation tensors. Note that \textit{Mimose} will reuse the previous checkpointing plans without introducing overhead when executing iterations of the same input sizes.

\begin{table}[htbp]
	\centering
	\footnotesize
	\setlength\tabcolsep{2pt}
	\setlength\extrarowheight{1pt}
	\caption{Overhead breakdown of \textit{Mimose} under 6 GB memory budget when executing one epoch, where the total overhead is normalized to the single-iteration time.}
	\label{table:overhead-breakdown}
	\begin{tabular}{ c|c|c|c }
		\hline
		Task & Collector & Estimator \& Scheduler & Total \\
		\hline
		\tabincell{c}{MC-Roberta \\(371.86 ms/iter)} & \tabincell{c}{145.99 ms \\(10 times)} & \tabincell{c}{0.26 ms\textasciitilde0.32 ms\\(17 times)} & \tabincell{c}{1464.77 ms \\(3.93 iters)}\\
		\hline
		\tabincell{c}{QA-XLNet \\(1033.56 ms/iter)} & \tabincell{c}{294.47 ms \\(10 times)} & \tabincell{c}{0.31 ms\textasciitilde0.39 ms\\(69 times)} & \tabincell{c}{2967.76 ms \\(2.87 iters)}\\
		\hline
		\tabincell{c}{QA-Bert \\(452.89 ms/iter)} & \tabincell{c}{118.96 ms \\(10 times)} & \tabincell{c}{0.27 ms\textasciitilde0.38 ms\\(24 times)} & \tabincell{c}{1196.88 ms \\(2.64 iters)}\\
		\hline
		\tabincell{c}{TC-Bert \\(250.27 ms/iter)} & \tabincell{c}{157.73 ms \\(10 times)} & \tabincell{c}{0.27 ms\textasciitilde0.34 ms\\(51 times)} & \tabincell{c}{1592.49 ms \\(6.36 iters)}\\
		\hline
	\end{tabular}
\end{table}

It can be observed that the data collector accounts for about 25\%\textasciitilde65\% of the time within a single iteration due to forwarding twice and the slower first iteration. To this end, we control to collect data by forwarding twice only in the first 10 iterations, and predict memory usage by memory estimator if necessary in the remaining iterations. In contrast, the overhead of memory estimator and scheduler is less than 1 ms, which is negligible compared to the single-iteration time (less than 0.2\%). Since similar input sizes share the same plan, the memory scheduler only needs to generate the checkpointing plan dozens of times during the entire epoch. The total overhead of \textit{Mimose} is only 3.95 iterations on average, whereas the training of one epoch contains thousands of iterations. In sum, it is effective to improve performance under memory budgets by using \textit{Mimose} to generate checkpointing plans for varying input sizes.

\subsection{Memory Consumption}

\begin{figure}
	\centering
	\subfloat[Multiple Choice]{
		{\includegraphics[width=0.49\linewidth]{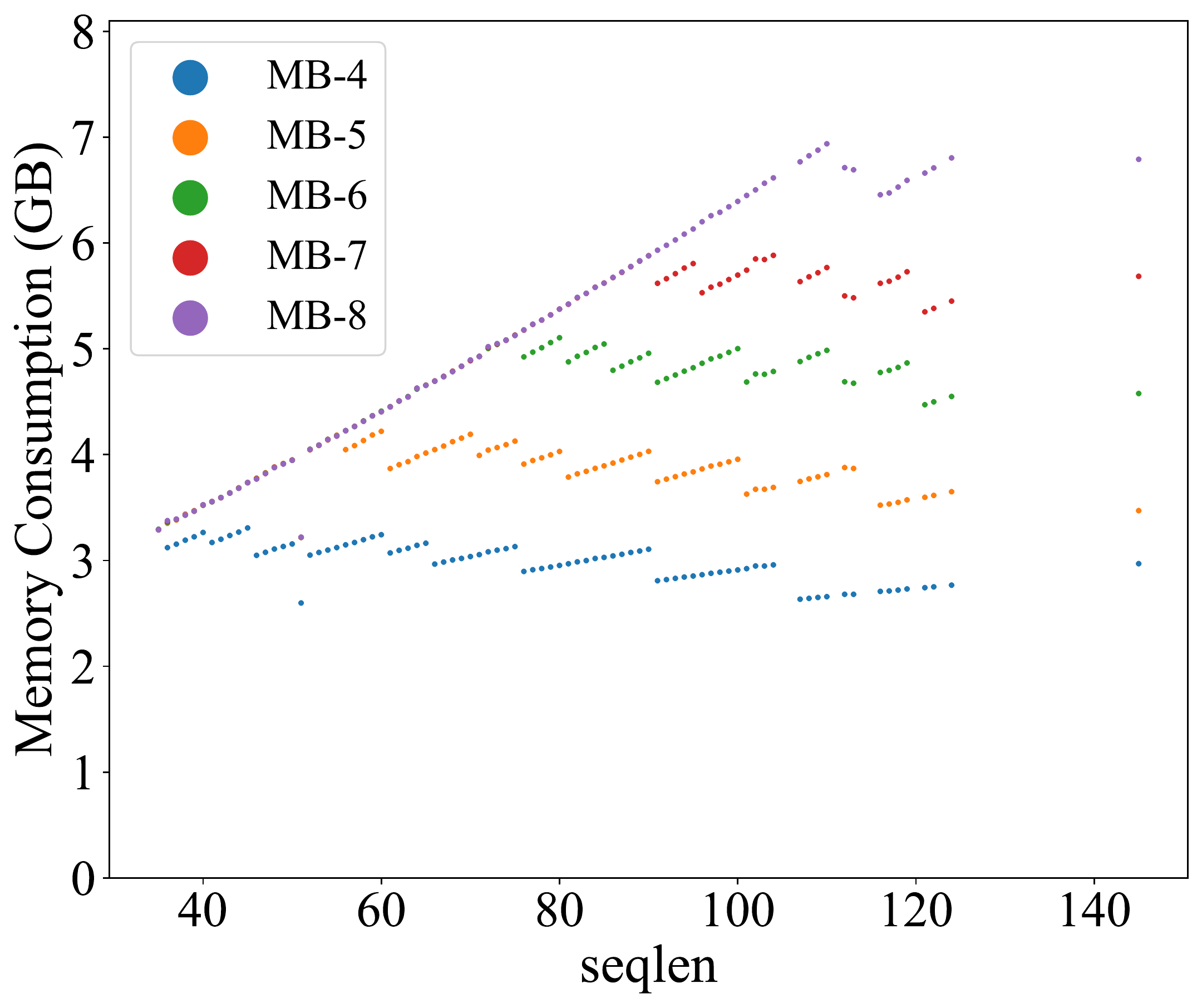} }
	}
	\subfloat[Question Answering (XLNet)]{
		{\includegraphics[width=0.49\linewidth]{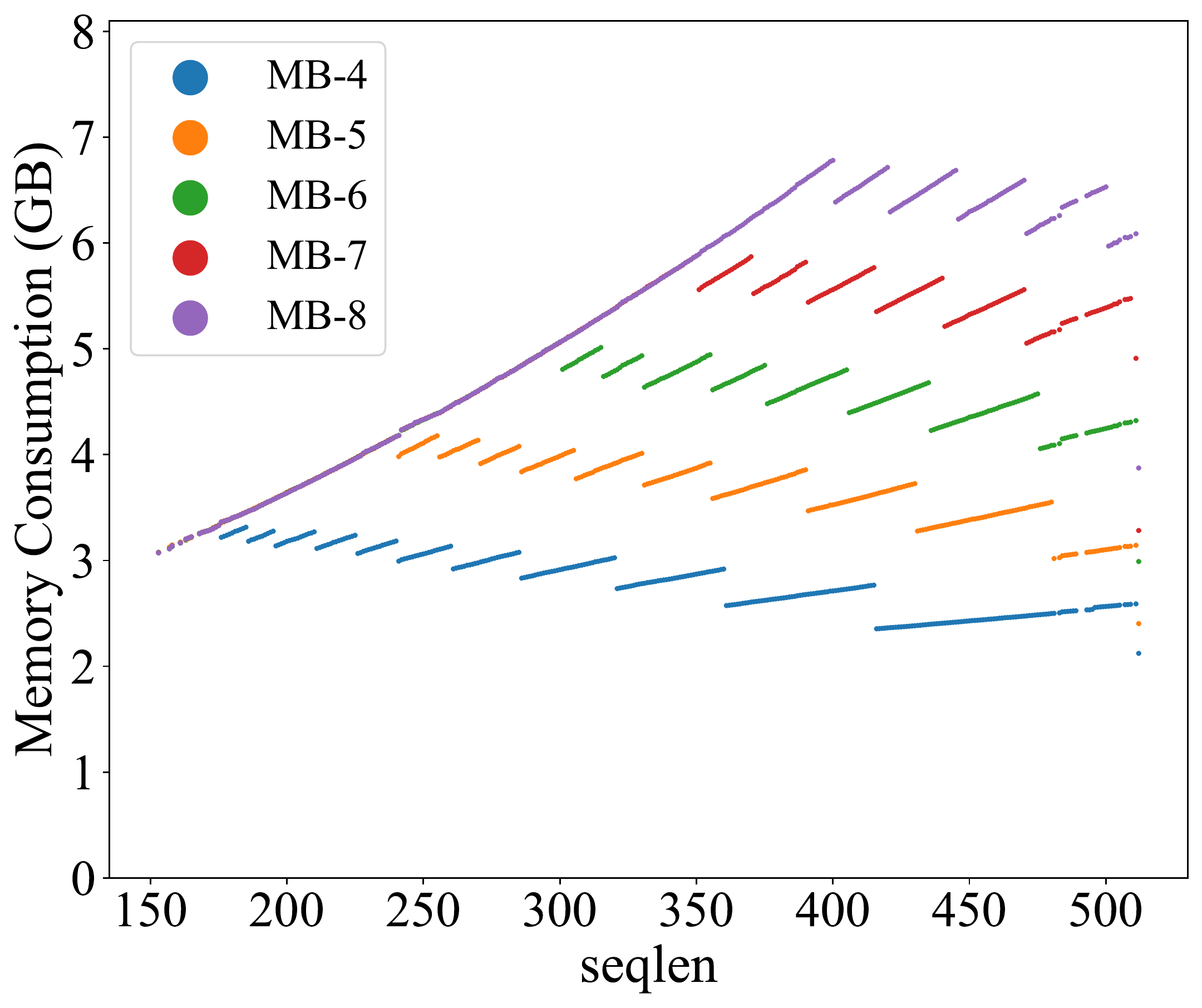} }
	}
	\vspace{-0.1in}
	\subfloat[Question Answering (Bert)]{
		{\includegraphics[width=0.49\linewidth]{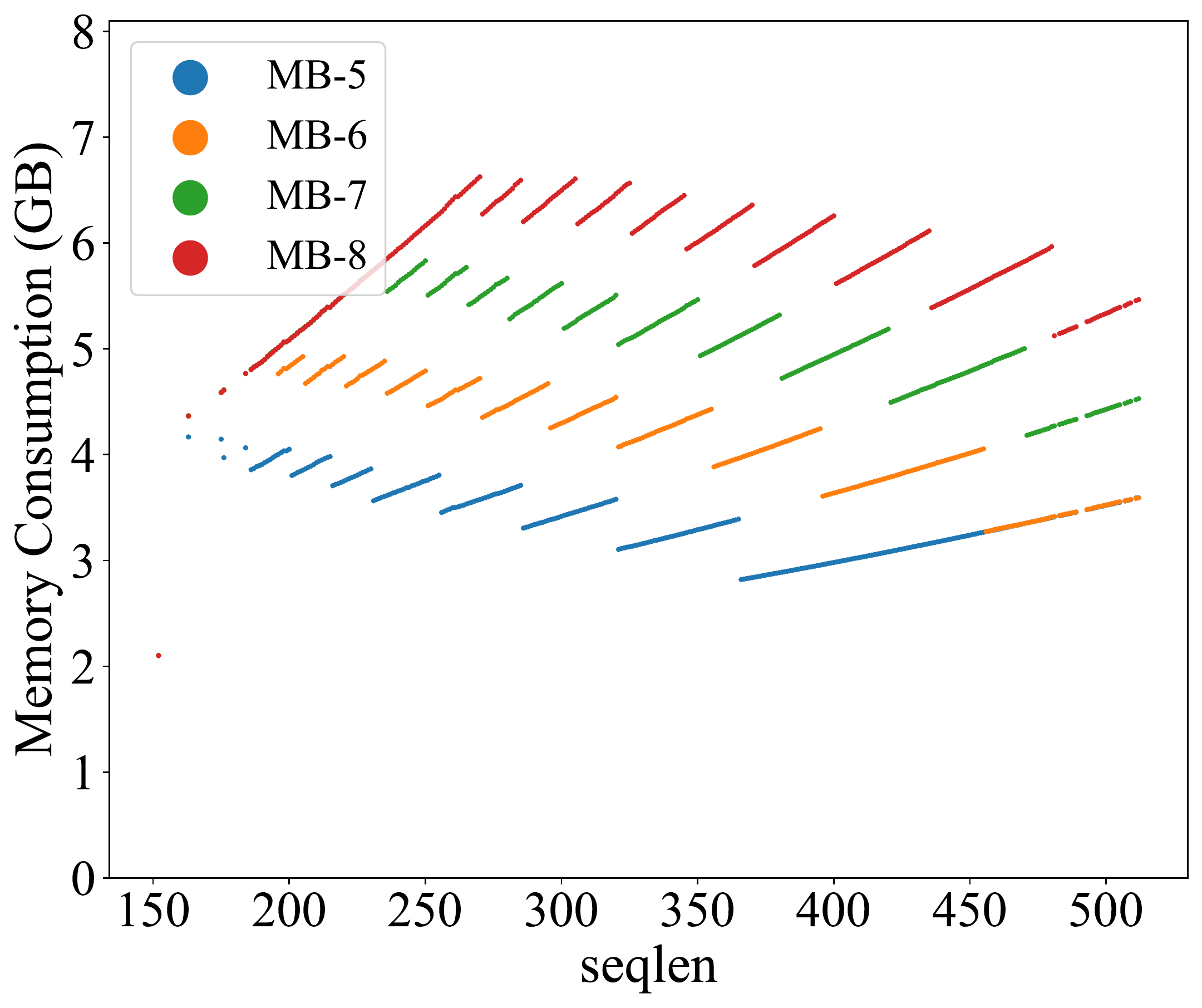} }
	}
	\subfloat[Text Classfication]{
		{\includegraphics[width=0.49\linewidth]{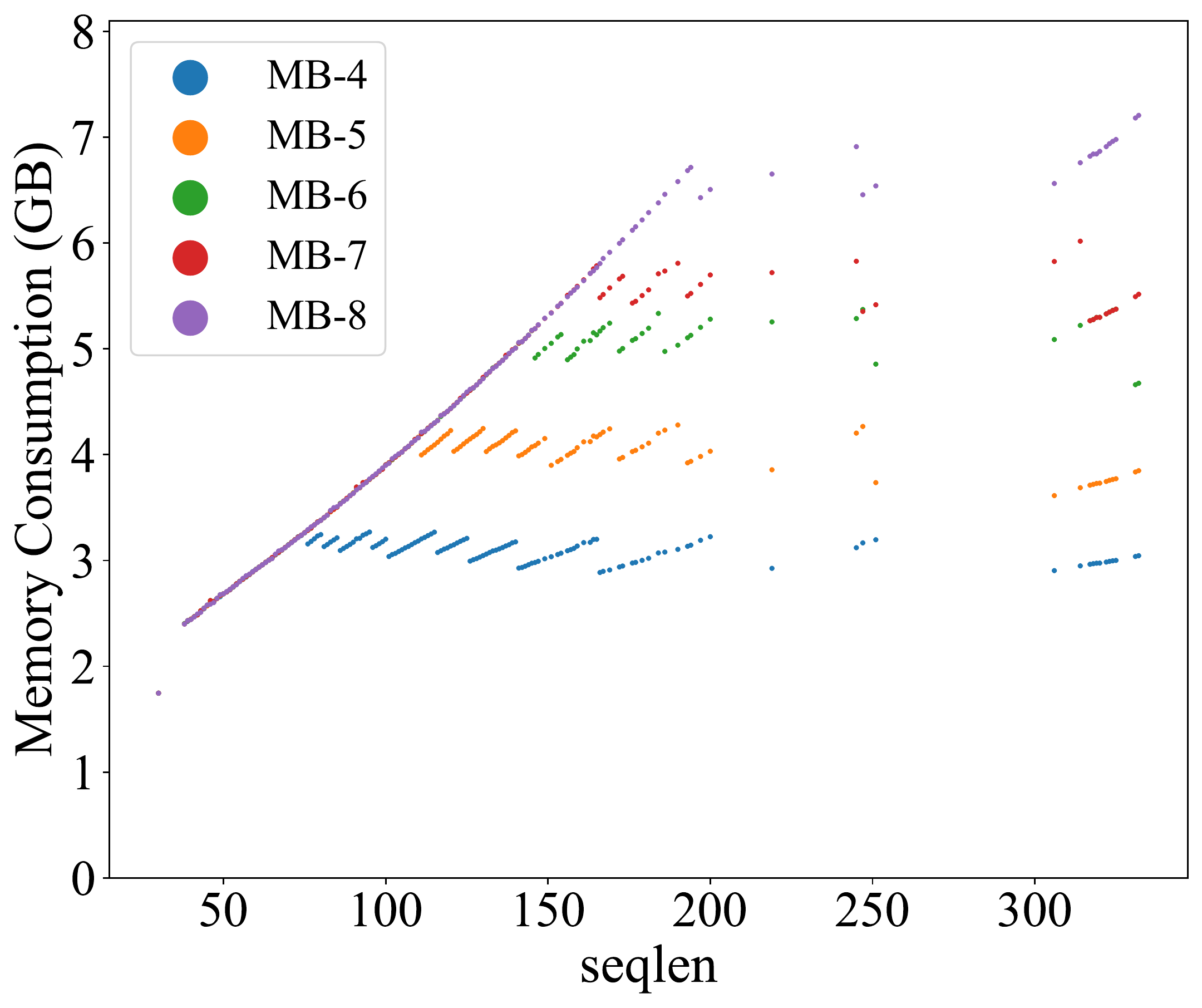} }
	}
	
	\caption{The memory consumption of \textit{Mimose} processing varying sequence lengths, where MB-\textit{X} refers to the memory budget of \textit{X} GB.}%
	\label{fig:memory-consumption}%
\end{figure}

\textit{Mimose} will adjust the checkpointing plan with the input size to minimize computational overhead. Figure~\ref{fig:memory-consumption} shows the memory consumption of \textit{Mimose} processing varying sequence lengths, where MB-\textit{X} refers to the memory budget of \textit{X} GB. It can be observed that there is a small gap between the upper limit of memory consumption and the memory budget. This is because \textit{Mimose} only needs to reserve 0.5 GB\textasciitilde1 GB of memory space to deal with possible memory fragmentation. In addition, there are a small number of points with particularly low memory consumption. The reason is that the data collector recomputes all modules in the first few iterations of the epoch to obtain layer-by-layer memory usage. In the next iterations, \textit{Mimose} avoids redundant recomputations by predicting memory usage through the estimation model.

The memory consumption increases with the input size until the memory budget is reached. This indicates that for small input sizes, memory optimization is disabled to avoid introducing redundant computations. After reaching the memory budget, \textit{Mimose} drops partial activation tensors through the checkpointing plan to reduce memory consumption. Since similar input sizes share the same checkpointing plan, the curve shows an upward trend of small segment separation with increasing input size. In addition, the curve of the latter segments under the same memory budget shows a downward trend. Consistent with other works~\cite{chen2016training,dtr}, the minimum recomputation unit of \textit{Mimose} is a layer (or module in other literature), and the memory consumption of each layer is positively related to the input size. The above results demonstrate that \textit{Mimose} can effectively utilize the memory budget by reducing memory fragmentation, thereby generating a near-optimal checkpoint plan with low computational overhead.

\subsection{Memory Prediction}
\label{sec:exp-prediction-model}
The memory estimator in \textit{Mimose} can be formulated as predicting the memory usage of the model under the given input size. 
We evaluate six representative regression models as candidates for our memory estimator, including the polynomial regression model(with order $n=1,2,3$), support vector machine (SVM), decision tree, and XGBoost. 
Specifically, we train the models using samples (i.e., per-layer memory usage under different input sizes) collected by the data collector, and compare the overall training time, prediction latency, and prediction error.
The experimental results on TC-Bert are shown in Table~\ref{table:memory-predictor-comparasion}. 
Except XGBoost, the training and prediction of the memory estimator candidates can be regarded as nearly zero-overhead compared to the per-iteration training time. 
For polynomial regression models with different orders($n=1,2,3$), the quadratic model achieves a very low prediction error (at the thousandth level), which demonstrates the correlation between the input sizes and the memory usages conforms to the quadratic polynomial distribution. 
Other regression models fail to achieve the same level of prediction error, even with more training samples, due to their tendency to overfit. 
Furthermore, the experiment results of the quadratic polynomial model on four training tasks are shown in Table~\ref{table:memory-predictor-polynomial}. The quadratic polynomial model achieved low prediction errors (at the thousandth level) on all tasks, confirming that our observation in Section~\ref{sec:memory-prediction} can be well generalized to NLP tasks. Therefore, we adopt the quadratic polynomial regression models in our memory estimator, which are more lightweight and more accurate than other regression models.

\begin{table}[htbp]
	\centering
	\footnotesize
	\setlength\tabcolsep{2pt}
	\caption{Prediction performance comparison of regression models on the text classification task, TC-Bert.}
	\label{table:memory-predictor-comparasion}
	\begin{tabular}{ c|c|c|c|c }
		\hline
		\tabincell{c}{Regression \\Model} & \# Samples & \tabincell{c}{Training Time \\(ms)} & \tabincell{c}{Prediction Latency \\(us)} & Error \\
		\hline
		Polynomial (n=1) & 10 & 0.90 & 14.78 & 4.04\% \\
		\textbf{Polynomial (n=2)} & \textbf{10} & \textbf{0.98} & \textbf{16.21} & \textbf{0.32}\% \\
		Polynomial (n=3) & 10 & 1.01 & 17.88 & 0.32\% \\
		SVR & 10 & 1.01 & 107.05 & 3.80\%\\
		SVR & 50 & 2.70 & 110.39 & 3.56\%\\
		DecisionTree & 10 & 3.98 & 82.97 & 5.67\%\\
		DecisionTree & 50 & 21.15 & 82.25 & 1.50\%\\
		XGBoost & 10 & 428.76 & 1348.26 & 5.13\%\\
		XGBoost & 50 & 2504.11 & 1354.93 & 1.43\%\\
		\hline
	\end{tabular}
\end{table}

\begin{table}[htbp]
	\centering
	\footnotesize
	\setlength\tabcolsep{2pt}
	\caption{Prediction performance of quadratic polynomial predictor on four training tasks.}
	\label{table:memory-predictor-polynomial}
	\begin{tabular}{ c|c|c|c|c }
		\hline
		Task & \# Samples & \tabincell{c}{Training Time \\(ms)} & \tabincell{c}{Prediction Latency \\(us)} & Error \\
		\hline
		MC-Roberta & 10 & 0.94 & 15.50 & 0.46\% \\ 
		QA-XLNet & 10 & 1.02 & 16.93 & 0.33\% \\
		QA-Bert & 10 & 1.18 & 16.45 & 0.33\% \\
		TC-Bert & 10 & 0.98 & 16.21 & 0.32\%\\
		\hline
	\end{tabular}
\end{table}

\subsection{Convergence}
\begin{figure}
	\centering
	\subfloat[Multiple Choice]{
		{\includegraphics[width=0.49\linewidth]{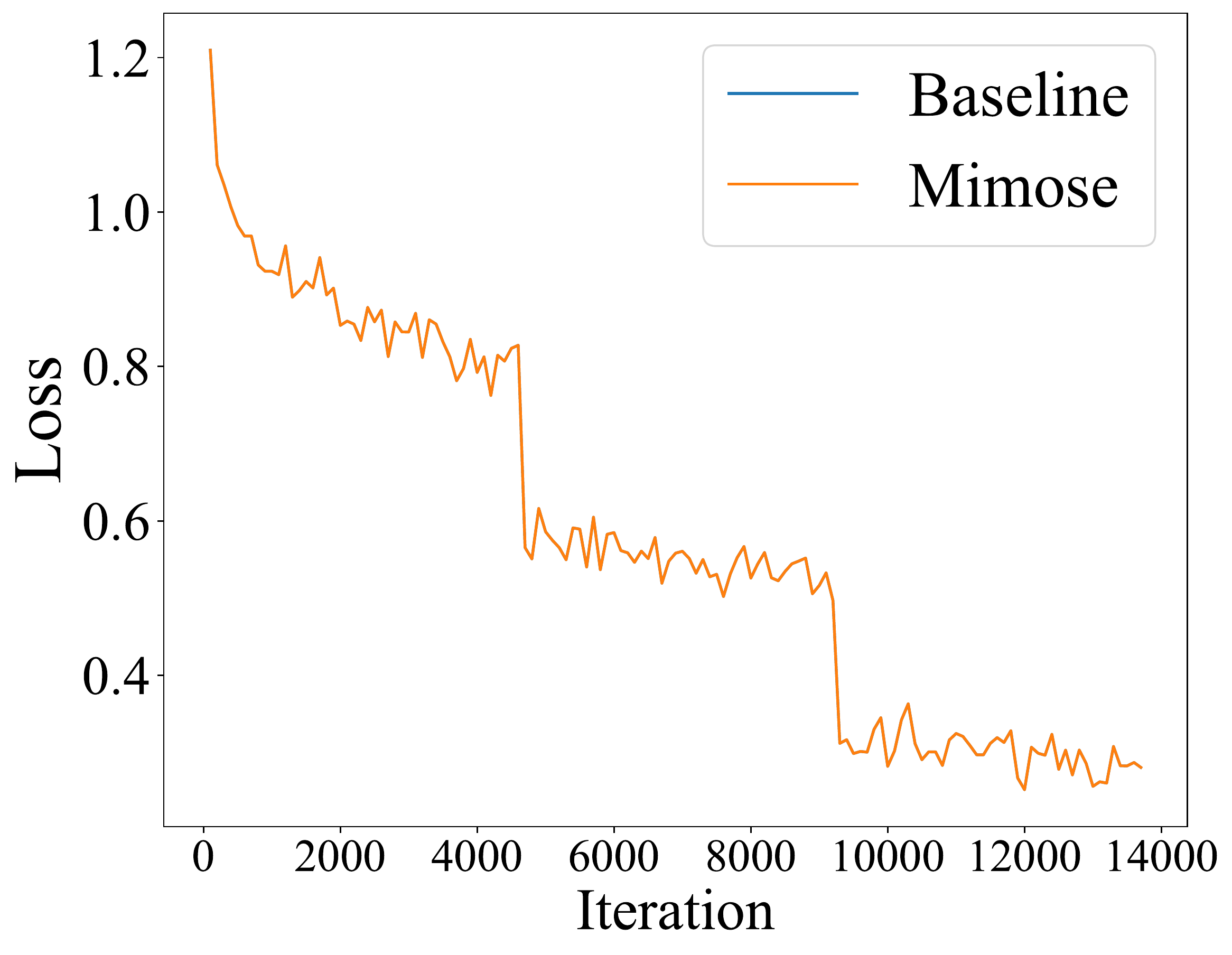} 
			\label{fig:loss-mc-roberta}
		}
	}
	\subfloat[Question Answering (XLNet)]{
		{\includegraphics[width=0.49\linewidth]{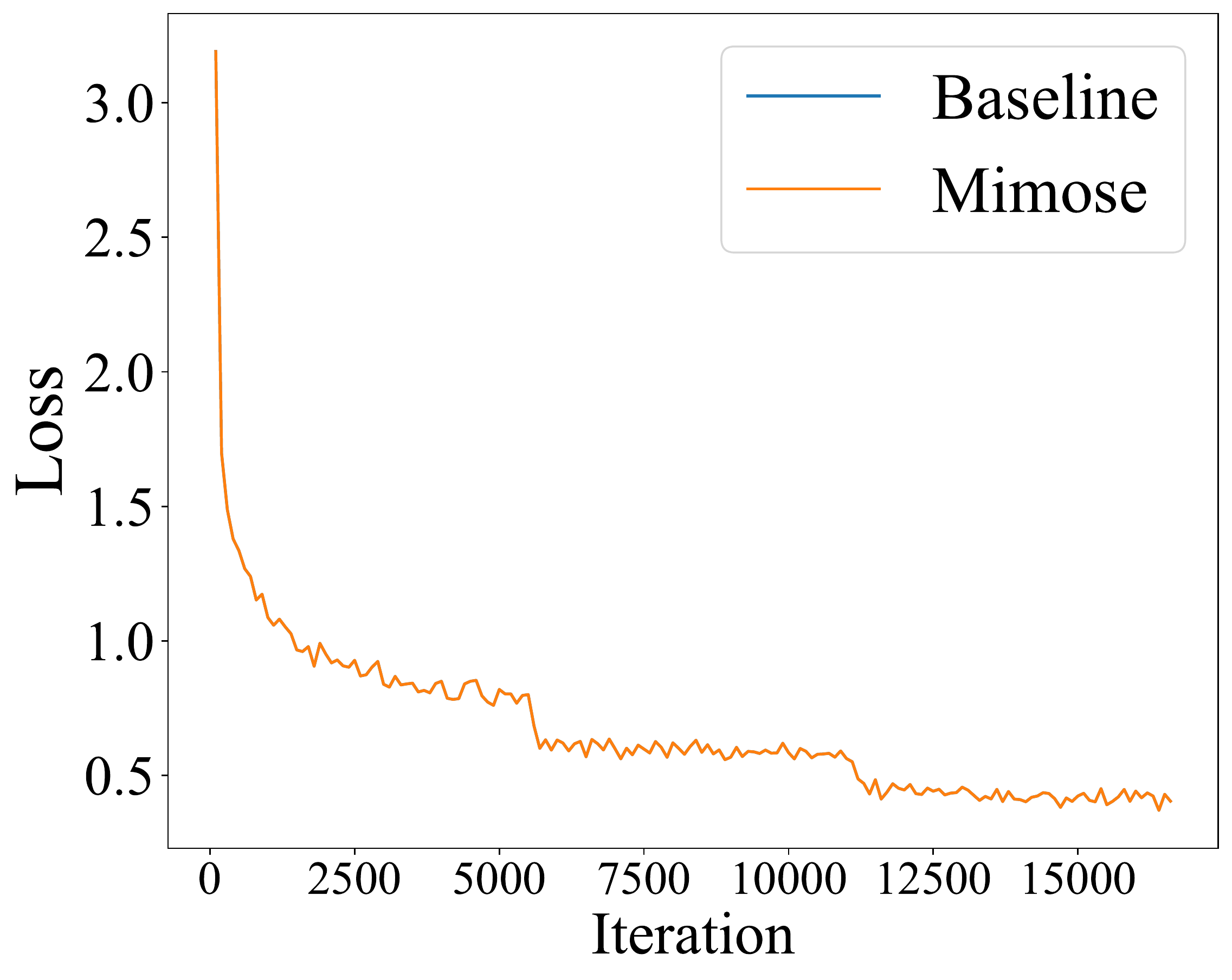}
			\label{fig:loss-qa-xlnet}
		}
	}
	\vspace{-0.1in}
	\subfloat[Question Answering (Bert)]{
		{\includegraphics[width=0.49\linewidth]{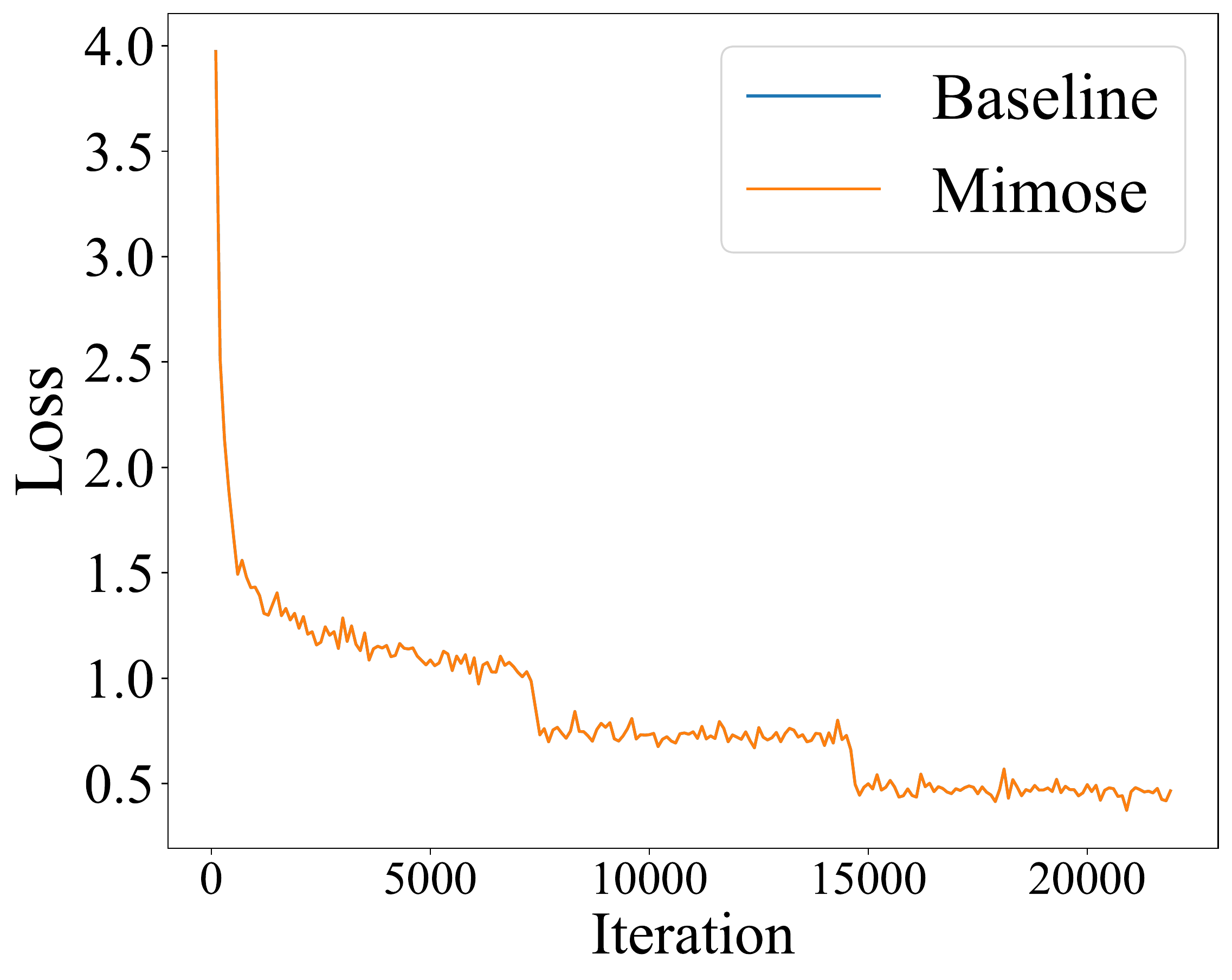} 
			\label{fig:loss-qa-bert}
		}
	}
	\subfloat[Text Classification]{
		{\includegraphics[width=0.49\linewidth]{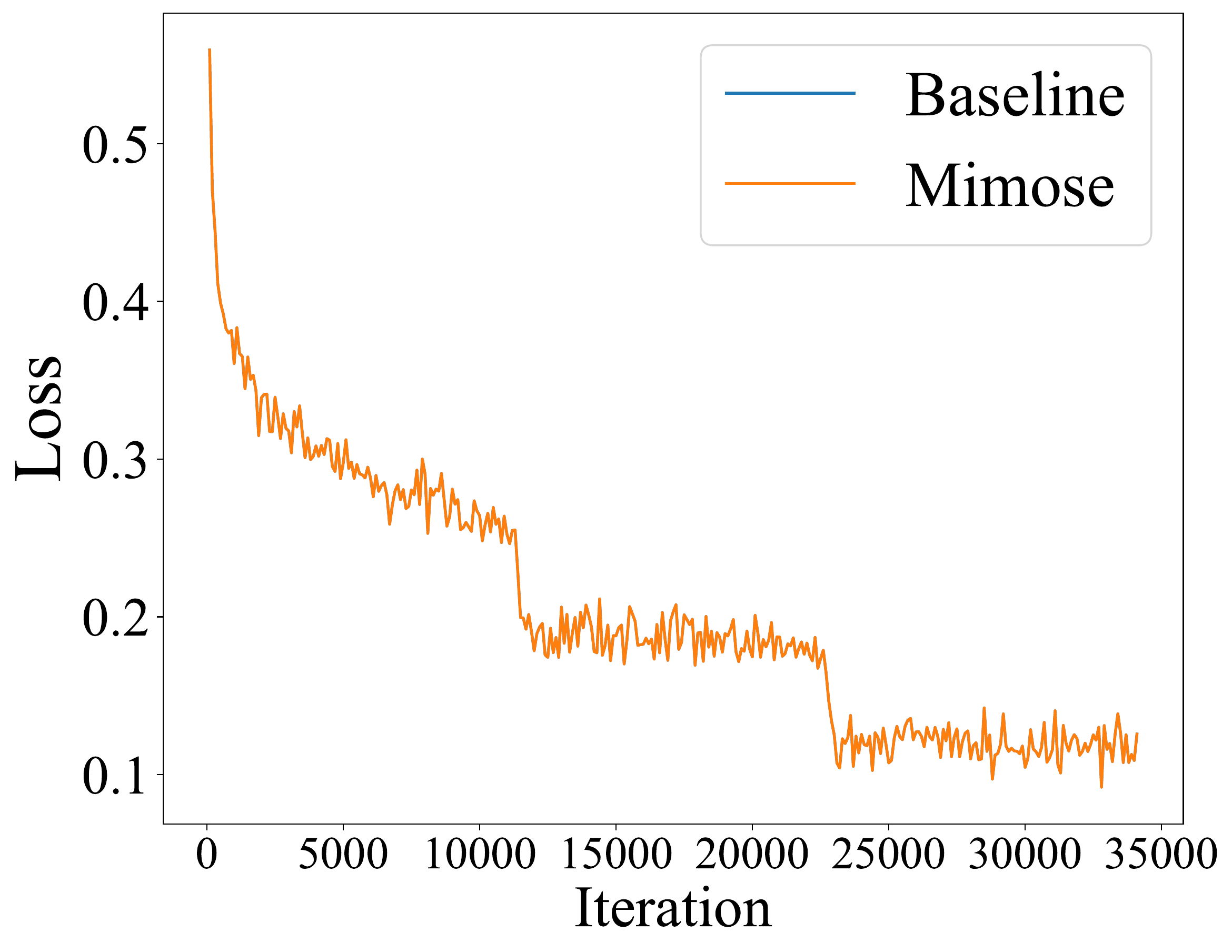}
			\label{fig:loss-tc-bert}
		}
	}
	\caption{Loss curves of training different tasks for three epoches.}%
	\label{fig:loss}%
\end{figure}

The training loss indicates the degree of prediction deviation from the true value. Figure~\ref{fig:loss} shows the loss curves of training different tasks for three epochs. We adopt the default configuration for initial learning rates (i.e., 5e-05, 3e-05, 3e-05, and 2e-05 for MC-Roberta, QA-XLNet, QA-Bert, and TC-Bert). \textit{Baseline} takes original PyTorch with no memory limit, whereas \textit{Mimose}'s memory budget is set to 6 GB. We skip the warm-up phase and start recording the loss values from the first 100 iterations. It can be observed that the loss of \textit{Mimose} gradually converges to an almost constant value. Furthermore, the curves of \textit{Mimose} and \textit{Baseline} are nearly coincident, which indicates that the modification of the computational graph by \textit{Mimose} does not affect the convergence. 

Note that the data collector obtains model information by forwarding twice in the first few iterations. At this time, the output results of layers (e.g., \texttt{dropout}) may be inconsistent due to the different states of the random number generator (RNG). To this end, \textit{Mimose} ensures that the output of each iteration is consistent with the normal forward execution by saving and restoring the RNG states.

\section{Related Work}
\label{sec:relatedwork}

\noindent \textbf{Model Compression for DNNs -} Since over-parameterization is a common property of DNN models, research works exploit compression techniques such as low precision~\cite{courbariaux2014training,gupta2015deep,judd2016proteus,jain2018gist}, quantization~\cite{han2015deep,hubara2016binarized}, and pruning~\cite{han2015learning,han2016eie,yang2021auto} to reduce memory consumption. For example, \textit{Courbariaux et al.}~\cite{courbariaux2014training} found that low precision multipliers were sufficient for training DNNs. \textit{Gupta et al.}~\cite{gupta2015deep} proved that DNNs could be trained using low-precision fixed-point arithmetric with stochastic rounding. \textit{Proteus}~\cite{judd2016proteus} adopted layer-specific precision representation for both neurons and weights to reduce the data traffic and storage footprint. \textit{Gist}~\cite{jain2018gist} reduced value redundancy in DNN training by storing the encoded representation of feature maps and decoding the data in the backward pass. 

Quantization and pruning techniques are also widely used to reduce network redundancy. \textit{Hubara et al.}~\cite{hubara2016binarized} introduced binarized neural networks (BNNs) that replaced most arithmetic operations with bit-wise operations. \textit{Han et al.}~\cite{han2015deep} quantized the weights to enforce weight sharing where only the effective weights and indices were stored. \textit{EIE}~\cite{han2016eie} leveraged sparsity in both the activations and weights after pruning the redundant connections to save the energy. \textit{AUTO-PRUNE}~\cite{yang2021auto} exploited reinforcement learning to search the pruning ratio of each layer considering the constraint of accuracy loss. However, compression techniques have an unpredictable impact on the convergence speed, and often requires customized hardware designs to ensure computational efficiency.



\noindent \textbf{Swapping with Recomputation -} Swapping techniques~\cite{rhu2016vdnn,meng2017training,shriram2019dynamic,huang2020swapadvisor,ren2021sentinel} expand the scale of DNN training under limited memory capacity by offloading temporarily unneeded data to the CPU. \textit{vDNN}~\cite{rhu2016vdnn} performed data swapping at layer granularity, where data was offloaded in forward phase and prefetched at backpropagation. \textit{Shriram et al.}~\cite{shriram2019dynamic} extended \textit{vDNN} that applied non-offload high-end allocation to reduce memory fragments. \textit{SwapAdvisor}~\cite{huang2020swapadvisor} used genetic algorithm to explore swapping decisions and designed data engine simulator to estimate the execution time. \textit{Sentinel}~\cite{ren2021sentinel} coordinated OS and runtime
profiling to offload tensors with similar access frequency into the same pages.

There are also research works~\cite{wang2018superneurons,jiang2019layup,capuchin,wen2022swap,he2022home,hu2022megtaichi} combining swapping with recomputation for hybrid memory optimization. \textit{SuperNeurous}~\cite{wang2018superneurons} introduced unified tensor pool to reuse heterogeneous memory and recomputed less expensive layers at back-propagation. \textit{Capuchin}~\cite{capuchin} performed mini-batches to obtain tensor access patterns and decoupled swapping with computation for synchronization reduction. \textit{STR}~\cite{wen2022swap} utilized mixed integer linear programming to generate the optimal execution plan that made full use of the swapped tensors. \textit{HOME}~\cite{he2022home} used the particle swarm algorithm with holistic model information to make tensor placement decisions. However, swapping techniques may achieve high copy overhead due to limited PCIe bandwidth. Especially for varying input tensors, it is difficult to dynamically adjust swapping decisions to hide latencies by overlapping.






\section{Conclusion}
\label{sec:conclusion}

In this paper, we propose \textit{Mimose}, an input-aware checkpointing planner for efficient GPU training under memory budgets. 
\textit{Mimose} is designed to be dynamic and agile in adjusting the checkpointing plan according to the predicted memory usage of the current input tensor to maximize GPU memory utilization and minimize the performance overhead introduced by checkpointing. 
\textit{Mimose} builds an online memory estimator to derive the activation tensor sizes with a given input tensor, whose data collection and estimation model are tailored and simplified based on identified memory usage patterns. 
Then, \textit{Mimose} exploits a responsive scheduler that generates checkpointing plans with respect to the cost-efficient layers and enables the fast switching of plans.
The experiment results show that \textit{Mimose} can achieve a shorter training time for training tasks from various NLP tasks. Although lying in the critical path of training, \textit{Mimose} itself only introduces almost negligible overhead, which is 3.95 iterations on average.

\bibliographystyle{plain}
\bibliography{references}

\end{document}